\documentclass[journal]{IEEEtran}

\usepackage{amssymb, latexsym, mathrsfs}
\usepackage{amsmath}
\usepackage{amsthm}
\usepackage{mathtools}
\usepackage{nicefrac}							
\usepackage{siunitx}							
\DeclareSIUnit{\pu}{pu}
\DeclareSIUnit{\var}{var}

\usepackage{bm}									
\usepackage{dsfont}
\usepackage{enumerate}
\usepackage{algorithm}
\usepackage{color}
\usepackage{transparent}
\usepackage{multicol}
\usepackage{booktabs}
\usepackage{etex}
\usepackage{shellesc}
\usepackage{subcaption}

\usepackage{graphicx}
\usepackage{xcolor}

\usepackage{pgfplots}
\pgfplotsset{compat=newest}
\usetikzlibrary{plotmarks}
\usetikzlibrary{arrows.meta}
\usepgfplotslibrary{patchplots}
\usepackage{grffile}

\usepackage{tikz}
\usepackage[american]{circuitikz}

\makeatletter
\let\NAT@parse\undefined
\makeatother
\usepackage[bookmarks=false]{hyperref}

\usepackage{indentfirst}
\pgfplotsset{plot coordinates/math parser=false}
\usetikzlibrary{positioning}

\usetikzlibrary{shapes,arrows}
\usetikzlibrary{calc}
\usetikzlibrary{angles}
\usetikzlibrary{positioning}
\usetikzlibrary{fit}
\usetikzlibrary{backgrounds}

\usepackage{arydshln}					

\newtheorem{theorem}{Theorem}
\newtheorem{assumption}{Assumption}
\newtheorem{problem}{Problem}

\newtheorem{remark}{Remark}
\newtheorem{proposition}{Proposition}
\newtheorem{lemma}{Lemma}
\newtheorem{definition}{Definition}

\IEEEoverridecommandlockouts

\newcommand{\CC}{{\mathcal{C}}}
\newcommand{\DD}{{\mathcal{D}}}
\newcommand{\EE}{{\mathcal{E}}}

\newcommand{\GG}{{\mathcal{G}}}

\newcommand{\KK}{{\mathcal{K}}}
\newcommand{\LL}{{\mathcal{L}}}
\newcommand{\MM}{{\mathcal{M}}}
\newcommand{\NN}{{\mathcal{N}}}

\newcommand{\PP}{{\mathcal{P}}}
\newcommand{\QQ}{{\mathcal{Q}}}
\newcommand{\RR}{{\mathcal{R}}}

\newcommand{\VV}{{\mathcal{V}}}

\newcommand{\XX}{{\mathcal{X}}}

	\definecolor{purpleDark}{RGB}{118, 4, 205}
	\definecolor{purpleLight}{RGB}{186, 102, 250}
	
	\definecolor{blueDark}{RGB}{52, 78, 243}
	\definecolor{blueLight}{RGB}{118, 135, 244}
	\definecolor{redDark}{RGB}{197, 34, 0}
	\definecolor{redLight}{RGB}{255, 91, 57}
	\definecolor{yellowDark}{RGB}{255, 183, 0}
	\definecolor{yellowLight}{RGB}{255, 204, 77}
	\definecolor{greenDark}{RGB}{0, 143, 53}
	\definecolor{greenLight}{RGB}{42, 189, 97}
	
	\colorlet{greenFaint}{green!10!white}
	\colorlet{redFaint}{red!10!white}

    \def\colorLine{blueDark}	
    
\definecolor{colorDGU1}{rgb}{0.00000,0.24000,0.72000}
\definecolor{colorDGU2}{rgb}{0.85000,0.32500,0.09800}
\definecolor{colorDGU3}{rgb}{0.92900,0.69400,0.12500}
\definecolor{colorDGU4}{rgb}{0.49400,0.18400,0.55600}
\definecolor{colorDGU5}{rgb}{0.00000,0.61000,0.52870}
\definecolor{colorDGU6}{rgb}{0,0,0}

	\newcommand{\dq}{\emph{dq} }				

	\newcommand{\addWithPreComma}[1]{%
		\if\relax #1\relax
		\else%
		,#1
		\fi%
	}
	\newcommand{\addWithPostComma}[1]{%
		\if\relax #1\relax
		\else%
		#1,
		\fi%
	}
	\newcommand{\addInParentheses}[1]{%
	\if\relax #1\relax
	\else%
		(#1)
	\fi%
	}

	\renewcommand{\vec}[1]{\bm{#1}}				
	\renewcommand{\matrix}[1]{\bm{#1}}		
	
	\newcommand{\Reals}{\mathbb{R}}		
	\newcommand{\Nats}{\mathbb{N}}		

	\newcommand{\Transpose}{\textsf{T}}	
	
	\newcommand{\Hamil}{H}

	\newcommand{\sDesired}{\text{c}}
	
	\newcommand{\sD}{\text{d}}
	\newcommand{\sQ}{\text{q}}
	\newcommand{\sDQ}{\text{dq}}
	
	\newcommand{\sLoad}{j}
	\newcommand{\sLine}{l}
	\newcommand{\sDisturbance}{\text{Z}}
	\newcommand{\sMG}{\text{MG}}
	
	\newcommand{\sPHS}{\text{PHS}}
	\newcommand{\sPI}{\text{PI}}
	\newcommand{\sInput}{\text{t}}
	\newcommand{\sSaturated}{\text{sat}}
	
	\newcommand{\sActuated}{\alpha}
	\newcommand{\sUnactuated}{\nu}
	\newcommand{\sIA}{\text{IA}}
	\newcommand{\sI}{\text{I}}
	\newcommand{\sZ}{\mathrm{\varkappa}}
	\newcommand{\sExtended}{\text{e}}

	\newcommand{\x}{\vec{x}}
	\newcommand{\xRef}{\bar{\vec{x}}}
	\newcommand{\xdot}{\dot{\x}}
	\newcommand{\xtilde}{\widetilde{\vec{x}}}
	\newcommand{\xtildedot}{\dot{\widetilde{\vec{x}}}}
	\newcommand{\y}{\vec{y}}
	
	\newcommand{\vu}{\vec{u}}
	\newcommand{\vuRef}{\bar{\vec{u}}}
	\newcommand{\vd}{\vec{d}}
	
	\newcommand{\z}{\vec{z}}
	\newcommand{\zRef}{\bar{\vec{z}}}
	
	\newcommand{\one}{\vec{1}}
    \newcommand{\zero}{\vec{0}}
	
	\newcommand{\Zstate}{\varkappa}
	\newcommand{\Zstates}{\vec{\varkappa}}
	
	\newcommand{\A}{\matrix{A}}
	\newcommand{\B}{\matrix{B}}
	\newcommand{\C}{\matrix{C}}
	\newcommand{\D}{\matrix{D}}
	
	\newcommand{\G}{\matrix{G}}
	
	\newcommand{\J}{\matrix{J}}
	\newcommand{\K}{\matrix{K}}
	\newcommand{\M}{\matrix{M}}
	\newcommand{\Q}{\matrix{Q}}
	\newcommand{\R}{\matrix{R}}
	
	\newcommand{\One}{\matrix{1}}
    \newcommand{\Zero}{\matrix{0}}
	
    \newcommand{\xL}{\x_{\sLoad}}
    \newcommand{\xLRef}{\xRef_{\sLoad}}
    
	\newcommand{\JL}{\J_{\sLoad}}
	\newcommand{\KL}{\K_{\sLoad}}
	\newcommand{\RL}{\mathcal{R}_{\sLoad}}
	
	\newcommand{\Hd}{\Hamil_{\sDesired}}
	
	\newcommand{\Jd}{\J_{\sDesired}}
	\newcommand{\Rd}{\R_{\sDesired}}

    \newcommand{\Id}{I^\sD}
    \newcommand{\Iq}{I^\sQ}
    \newcommand{\Idq}{\vec{I}^\sDQ}

    \newcommand{\Vd}{V^\sD}
    \newcommand{\Vq}{V^\sQ}
    \newcommand{\Vdq}{\vec{V}^\sDQ}

     \newcommand{\Vdqdot}{\dot{\vec{V}}^{\sDQ}}
    
    \newcommand{\Ilined}{I_{\sLine}^{\sD}}
    
    \newcommand{\Ilineq}{I_{\sLine}^{\sQ}}
    
    \newcommand{\Ilinedq}{\vec{I}_{\sLine}^{\sDQ}}
    
    \newcommand{\Izd}[1][]{I^{\sD}_{\sDisturbance {#1}}}
    \newcommand{\Izq}[1][]{I^\sQ_{\sDisturbance {#1}}}
    \newcommand{\Izdq}[1][]{\vec{I}^\sDQ_{\sDisturbance {#1}}}
    
    \newcommand{\ILdq}{\vec{I}^\sDQ_{\sLoad }}
    
    \newcommand{\VLd}{V^\sD_{\sLoad }}
    \newcommand{\VLq}{V^\sQ_{\sLoad }}
    \newcommand{\VLdq}{\vec{V}^\sDQ_{\sLoad }}
    
    \newcommand{\IdRef}[1][]{\bar{I}^\sD_{\sInput{#1}}}
    \newcommand{\IqRef}[1][]{\bar{I}^\sQ_{\sInput{#1}}}
    \newcommand{\IdqRef}[1][]{\bar{\vec{I}}_{\sInput{#1}}^\sDQ}

    \newcommand{\VdRef}{V^{\sD *}}
    \newcommand{\VqRef}{V^{\sQ *}}
    \newcommand{\VdqRef}{\vec{V}^{\sDQ *}}
    
    \newcommand{\Itd}[1][]{I_{\sInput{#1}}^\sD}
    \newcommand{\Itq}[1][]{I_{\sInput{#1}}^\sQ}
    \newcommand{\Itdq}[1][]{\vec{I}_{\sInput{#1}}^\sDQ}

    \newcommand{\Vtd}[1][]{V^\sD_{\sInput{#1}}}
    \newcommand{\Vtq}[1][]{V^\sQ_{\sInput{#1}}}
    \newcommand{\Vtdq}[1][]{\vec{V}^\sDQ_{\sInput{#1}}}

    \newcommand{\Vamp}[1][]{V_{#1}}
    
    \newcommand{\VampNominal}{V_{0}}

	\newcommand{\freqRef}{\omega_0}
	
	\newcommand{\lAct}[1][]{P_\sLoad\addWithPreComma{#1}}
	\newcommand{\lReac}[1][]{Q_\sLoad\addWithPreComma{#1}}
	
	\newcommand{\lActNom}[1][]{P_{0\addWithPreComma{#1}}}
	\newcommand{\lReacNom}[1][]{Q_{0\addWithPreComma{#1}}}

	\newcommand{\lYp}[1][]{Y_{\text{P}{#1}}}
	\newcommand{\lIp}[1][]{I_{\text{P}{#1}}}
	\newcommand{\lPp}[1][]{P_{\text{P}{#1}}}

	\newcommand{\lYq}[1][]{Y_{\text{Q}{#1}}}
	\newcommand{\lIq}[1][]{I_{\text{Q}{#1}}}
	\newcommand{\lPq}[1][]{P_{\text{Q}{#1}}}
	
	\newcommand{\lNp}[1][]{n_{\text{P}{#1}}}
	\newcommand{\lNq}[1][]{n_{\text{Q}{#1}}}
	
	\newcommand{\Rt}{R_{\text{t}}}
	\newcommand{\Ct}{C_{\text{t}}}
	\newcommand{\Lt}{L_{\text{t}}}
	\newcommand{\Rti}{R_{\text{t}i}}
	\newcommand{\Cti}{C_{\text{t}i}}
	\newcommand{\Lti}{L_{\text{t}i}}
	
\begin{document}

\title{A Unified Passivity-Based Framework for Control of Modular Islanded AC Microgrids}

\author{Felix Strehle,
        Pulkit Nahata,
        Albertus Johannes Malan,
        S{\"o}ren Hohmann,
        Giancarlo Ferrari-Trecate
\thanks{This work was supported in part by the Swiss National Science Foundation through the COFLEX Project (Grant 200021\_169906) and NCCR Automation  (Grant 51NF40\_180545). (\textit{Corresponding author: Felix Strehle).}}
\thanks{F.\ Strehle, A.\ J.\ Malan, S.\ Hohmann are with the Institute of Control Systems (IRS), Karlsruhe Institute of Technology (KIT), Germany (e-mail: felix.strehle@kit.edu; albertus.malan@kit.edu; soeren.hohmann@kit.edu). }
\thanks{P.\ Nahata, and G.\ Ferrari-Trecate are with the Institute of Mechanical Engineering, \'Ecole Polytechnique F\'ed\'erale de Lausanne, Switzerland (e-mail: pulkit.nahata@epfl.ch; giancarlo.ferraritrecate@epfl.ch).}%
}

\maketitle
\begin{abstract}
Voltage and frequency control in an islanded AC microgrid (ImGs) amount to stabilizing an a priori unknown ImG equilibrium induced by loads and changes in topology. 
This paper puts forth a unified control framework which, while guaranteeing such stability, allows for modular ImGs interconnecting multiple subsystems, that is, dynamic RLC lines, nonlinear constant impedance, current, power (ZIP) and exponential (EXP) loads, and inverter-based distributed generation units (DGUs) controlled with different types of primary controllers.
The underlying idea of the framework is based on \emph{equilibrium-independent passivity} (EIP) of the ImG subsystems, which enables stability certificates of ImG equilibria without their explicit knowledge.
In order to render DGUs EIP, we propose a decentralized controller synthesis algorithm based on port-Hamiltonian systems (PHSs).
We also show that EIP, being the key to stability, provides a general framework which can embrace other solutions available in the literature. 
Furthermore, we provide a novel argument based on LaSalle's theorem for proving asymptotic voltage and frequency stability.
Finally, we analyze the impact of actuator saturation on the stability results by exploiting the inherent EIP properties of the PHS DGU model. 
Theoretical findings are backed up by realistic simulations
based on the CIGRE benchmark for medium voltage networks.
\end{abstract}


%
\IEEEpeerreviewmaketitle

\section{Introduction}
AC microgrids have been identified as a key element of future electrical supply systems. Operating from low to medium voltage levels, they provide a systemic, modular approach to coping with the time-varying topologies caused by a rising integration of flexible loads and DGUs \cite{Lasseter01, Schiffer16, Olivares14, Guerrero13, Farrokhabadi20ieee}. As DGUs commonly comprise intermittent renewable energy sources and storage devices operating in DC, they interface with the remaining AC microgrid via controllable DC-AC \emph{voltage source inverters} (VSIs) and RLC filters (see Fig.~\ref{fig:dgu}) \cite{Schiffer16, Olivares14, Guerrero13}.
 
 A major challenge in microgrid control is operation in \emph{islanded} mode, where the main grid no longer acts as an infinite power source and a master clock for the network frequency. As a result, in ImGs, basic voltage and frequency stabilization at primary level must be performed exclusively by DGUs via their local VSI controllers \cite{Olivares14, Guerrero13, Farrokhabadi20ieee}. At the same time, a large number of interacting subsystems (DGUs, loads), along with their intermittent supply/demand behavior calls for decentralized scalable primary controllers which, in the event of an ImG topology change, either remain unaffected or are easily updatable  \cite{Lasseter01}.
 Decentralized control design methods relying only on local information of a given DGU, ensure such scalability. They allow for the addition or removal of subsystems in a plug-and-play fashion, that is,  without adapting other controllers, communicating, and endangering voltage and frequency stability. 

Approaches to the decentralized control of ImGs can be divided into two main classes. The first one comprises droop control and extensions thereof (see for example \cite{Olivares14, Guerrero13, Doerfler14, Schiffer14conditions, Zhong13}. Despite its popularity, droop control shows load-dependent voltage and frequency deviations from the nominal values, propagation of voltage errors along resistive transmission lines, poor performance at distribution level, where a low $X/R$ ratio results in a non-negligible coupling of active and reactive power, and the presence of steady-state voltage drifts \cite{Olivares14, Guerrero13, Schiffer14conditions}. The compensation of these issues entails the use of distributed secondary controllers along with some form of communication \cite{Olivares14, Guerrero13, Kolluri18}. 
Stability properties of droop-controlled microgrids have been analyzed for example in \cite{Schiffer14conditions,SimpsonPorco13a, SimpsonPorco13b} under simplified, first-order models of the VSI dynamics, quasi steady-state network models, and simplified load models.
 
The second class of controllers comprises droop-free approaches which usually implement voltage controllers in the $\dq$ frame and use open-loop frequency control and GPS synchronization to stabilize system frequency \cite{Etemadi12a,Babazadeh13,Riverso15,Tucci17ac,Sadabadi17}. 
 Within this second class, several approaches are based on the concept of neutral interactions \cite{Riverso15, Tucci17ac,Sadabadi17}. However, they require quasi-stationary line approximations and necessitate solving linear matrix inequalities after each DGU plug-in/out. 
 In \cite{Nahata19ecc}, a passivity-based extension of \cite{Tucci17ac} has been proposed. 
 However, in both \cite{Tucci17ac} and \cite{Nahata19ecc}, the stability proof assumes linear loads. Furthermore, the controllers are restricted to a PI-control structure.
Despite their non-applicability to ImGs which feed nonlinear loads inducing unknown equilibria, results in \cite{Nahata19ecc} together with \cite{Strehle19} suggest that compositional properties of passive systems can provide a promising unifying control framework towards realizing modular ImGs in which topologies change frequently and different VSI controllers dynamically interact via the network (see also \cite{Fiaz13,vdS16, Gui18, Monshizadeh19}).

Furthermore, pertinent literature usually neglects the presence of actuator saturation naturally arising from the physical constraints of the inverters (see for example \cite{Schiffer14conditions, SimpsonPorco13a,SimpsonPorco13b,Etemadi12a,Babazadeh13,Sadabadi17,Riverso15,Tucci17ac, Cucuzzella18, Perez04, Zhong17, Serra17b,Tucci18TechRep}).
 
The contributions of this work are five-fold.
 \emph{First}, by availing ourselves of the innate skew symmetry of ImG interconnections \cite{Nahata19ecc}, we demonstrate that EIP\footnote{EIP is a stronger notion than mere passivity as it ensures passivity with respect to any feasible equilibrium and thus stability certificates without explicit knowledge of the equilibirium \cite{Hines11,Burger14,Arcak16}.} of the ImG subsystems is a sufficient condition for guaranteeing voltage and frequency stability, i.e.\ stability of any unknown ImG equilibrium induced by a change in topology or load fluctuations. This establishes a compositional, unifying control framework wherein multiple ImG subsystems, if EIP, can enter or leave the ImG network without having any bearing on stability properties. 
 \emph{Second}, unlike previous contributions \cite{Tucci17ac,Nahata19ecc} limited to either linear loads, static lines, or both, this work considers ImGs composed of controlled DGUs with RLC filters, dynamic RLC lines, and nonlinear ZIP and EXP loads. 
 \emph{Third}, we extend our port-Hamiltonian-based control design from \cite{Strehle19} by adding an integral action (IA). The result is a voltage controller in $\dq$ coordinates for DGU VSIs which is robust against parameter uncertainties and ensures zero voltage errors in the steady-state.
 Differently from \cite{Tucci17ac, Nahata19ecc}, the proposed framework for control design does not require to \emph{a priori} specify the control structure and then built candidate Lyapunov functions for certifying passivity of the closed-loop system. Instead, the control law follows naturally as a consequence of the control requirements and the physical, passivity-related insight of PHSs \cite{vdS17}.  
 Note that, similar to \cite{Etemadi12a,Babazadeh13,Riverso15,Tucci17ac, Cucuzzella18}, we control the frequency in open-loop and use GPS synchronization. Thus, our $\dq$ voltage controllers are able to simultaneously stabilize bus voltages and system frequency. 
 \emph{Fourth}, with the goal of consolidating voltage and frequency stability, we establish that (i) the proposed $dq$ voltage controller renders DGUs EIP; (ii) nonlinear ZIP and EXP loads are strictly EIP under sufficient conditions set out in Section \ref{sec:stability:eip_load}; and (iii) lines are inherently EIP. 
 In order to demonstrate how EIP allows for a general, unifying control framework, we also consider the controllers from \cite{Nahata19ecc} in our stability analysis and subsequent simulations.
 We show that these controllers render DGUs EIP and thus can be readily used along with those proposed in the work. However, any other DGU VSI controllers---for example the passivity-based designs for single inverters in \cite{Perez04, Zhong17, Serra17b, Bobtsov20}---or even other components such as flexible AC transmission systems (FACTs) could be used as long as the closed-loop system is EIP. Differently from \cite{Strehle19, Strehle20ifac}, we also provide a novel, comprehensive proof based on LaSalle's theorem for asymptotic voltage and frequency stability of a modular ImG.
\emph{Fifth}, in contrast to pertinent literature and previous works \cite{Nahata19ecc, Strehle19}, this paper analyzes the impact of actuator saturation on the asymptotic stability of the ImG equilibrium. For this, we leverage the inherent EIP properties of the open-loop PHS DGU model used in the control design.
Simulations based on a modified version of the CIGRE benchmark for medium voltage networks \cite{Strunz14} illustrate our theoretical findings.

 \subsection{Preliminaries and notation}
 
\subsubsection{Sets, vectors, and functions} We let $\mathbb{R}$ (resp. $\mathbb{R}_{>0}$) denote the set of real (resp. strictly positive real) numbers. Given $ \x \in \mathbb{R}^{n}$, $\text{Diag}(x) \in \mathbb{R}^{n \times n}$ is the associated diagonal matrix with $x$ on the diagonal. 
The notation $A \succ 0$ ($A \succeq 0$) represents a positive definite  (positive semidefinite) matrix or function. 
Throughout, 
$\zero_n$, $\one_n$ are $n$-dimensional vectors of zero and unit entry, whereas $\Zero_{n\times n}$, $\One_{n\times n}$ are $n\times n$-dimensional zero and identity matrices.
A specific known or calculated equilibrium is denoted by $(\cdot)^*$, whereas an arbitrary, unknown equilibrium is denoted by $\bar{(\cdot)}$. The double notation $\bar{(\cdot)}^*$ indicates that an equilibrium comprises both, known and unknown parts.
The superscript \dq denotes a vector of instantaneous $d$ and $q$ components in the \dq coordinate frame rotating at $\freqRef = 2 \pi \, \SI{50}{\hertz}$, i.e.\ $\Vdq := \left[\Vd, \Vq\right]^{\textsf{T}}$. 
Functions and matrices of the desired \emph{closed-loop} system are denoted with the subscript $\sDesired$.

\subsubsection{Algebraic graph theory} We denote by $\mathcal{G}(\mathcal{V},\mathcal{E})$ a digraph, where $\mathcal{V}=\{1,\cdots,N\}$ is the node set and $\mathcal{E} \subseteq (\mathcal{V}\times\mathcal{V})$ is the edge set. 
All digraphs in this work are assumed to be without self loops, that is, $(i,i) \notin \mathcal{E}$. For node $i \in \mathcal{V}$,  $\mathcal{N}^+_i=\{j \in \mathcal{V}: (i,j) \in \mathcal{E}\}$ denotes the set of out-neighbors, $\mathcal{N}^-_i=\{j \in \mathcal{V}: (j,i) \in \mathcal{E}\}$ the set of in-neighbors, and $\mathcal{N}_i=\mathcal{N}^+_i\cup\mathcal{N}^-_i$ the set of neighbors.

\subsubsection{Equilibrium-Independent Passivity \emph{(EIP)} \cite{Hines11, Burger14}\cite[p.~24]{Arcak16}} Consider an autonomous, composite system
\begin{subequations} \label{eq:prems:comp_system}
\begin{equation}
    \dot{\x}=\vec{f}(\x)
\end{equation}   
composed of $N \in \Nats$ subsystems
\begin{align} \label{eq:prems:subsystem}
\dot{\x_i} = \vec{f}_i(\x_i,\vd_i), \quad \z_i = \vec{h}_i(\x_i,\vd_i)
\end{align}
with states $\x_i \in \Reals^{n_i}$, interaction (coupling) input $\vd_i \in  \Reals^{m_i}$, interaction (coupling) output $\z_i \in \Reals^{m_i}$, and interconnection structure 
\begin{equation} \label{eq:prems:static_intercon}
    \vd=
    \begin{bmatrix}
    \vd_1\\
    \vdots\\
    \vd_N
    \end{bmatrix}\!\!= \M 
    \begin{bmatrix}
    \vec{h}_1(\x_1,\vd_1)\\
    \vdots\\
    \vec{h}_N(\x_N,\vd_N)
    \end{bmatrix}\!\!
    = \M 
    \begin{bmatrix}
    \z_1\\
    \vdots\\
    \z_N
    \end{bmatrix}\!\!=\M \z
\end{equation}
\end{subequations}
with the static interconnection matrix $\M \in \Reals^{m \times m}$. Assume \eqref{eq:prems:comp_system} to be well-posed, i.e.\ after substituting $\z_i = \vec{h}_i(\x_i,\vd_i), i=1,\dots,N$ in \eqref{eq:prems:static_intercon}, a unique solution can be found for $\vd$ as a function of $\x$ \cite[p.~13]{Arcak16}. Then, suppose there exists a set $\bar{\XX}_i \subset \Reals^{n_i}$ of feasible, but \emph{unknown} equilibria $\xRef_i \in \bar{\XX}_i$ for each subsystem \eqref{eq:prems:subsystem}. Associated to every  $\xRef_i$, there is a unique input $\bar{\vd}_i \in \Reals^m$ satisfying $\vec{0} = \vec{f}_i(\xRef_i,\bar{\vd}_i)$. Consequently, $\bar{\vd}_i$ and the corresponding output $\zRef_i = \vec{h}_i(\xRef_i,\bar{\vd}_i)$ are implicit functions of $\xRef_i$.
\begin{definition} \label{def:eip}  
Subsystem \eqref{eq:prems:subsystem} is \emph{EIP}\footnote{Particularly in the context of PHS, EIP is also termed \emph{shifted passivity} \cite[p.~169]{vdS17} or \emph{passivity of incremental systems} \cite{ Burger14, Jayawardhana07}. However, \emph{incremental passivity} as defined in \cite{Stan07} \cite[p.~94--95]{vdS17}, i.e.\ passivity with respect to two arbitrary state trajectories $\x_1(t), \x_2(t)$, is more restrictive than EIP, where one of the trajectories is an equilibrium $\xRef$ \cite{Hines11, SimpsonPorco19}.}, if a $\CC^1$ storage function $S_i: \Reals^{n_i}\times \bar{\XX}_i \to \Reals_{\geq 0} $ exists, which satisfies $\forall(\x_i,\xRef_i,\bar{\vd}_i) \in \Reals^{n_i} \times \bar{\XX}_i \times \Reals^{m_i}$
	\begin{subequations} \label{eq:prems:eip_conditions}
	\begin{align}                                                                            
	S_i(\x_i,\xRef_i) &\geq 0\,, \quad S_i(\xRef_i,\xRef_i) = 0 \label{eq:prems:def_eip_posdef}   \\
	\dot{S}_i(\x_i,\xRef_i) &\leq (\vd_i-\bar{\vd}_i)^\Transpose(\z_i-\zRef_i)\,. \label{eq:prems:def_eip_neg_semidef}
	\end{align}
	\end{subequations}
If \eqref{eq:prems:def_eip_neg_semidef} is a strict inequality, \eqref{eq:prems:subsystem} is called \textit{strictly EIP}.
\end{definition}
While proving EIP is in general a challenging task for an arbitrary, nonlinear subsystem \eqref{eq:prems:subsystem}, there exist classes of systems which are inherently EIP.
\begin{lemma}\label{lemma:eip_systems}
	EIP is ensured for the following system classes:
	\begin{enumerate}
		\item[I)] linear, passive systems
		\begin{align}\label{eq:linear_sys}
		\dot{\vec{x}}_i = \A_i\x_i+\B_i\vd_i, \qquad \z_i= \C_i\vec{x}_i + \D_i\vd_i
			\end{align} 
		with the quadratic, positive definite storage function $	S_i(\x_i)=\frac{1}{2}\x_i^\Transpose \Q_i \x_i, \Q_i \succ 0$ \cite[p.~11]{Stan07} \cite[p.~26]{Arcak16}.
		\item[II)] linear \emph{input-state-output port-Hamiltonian systems} (ISO-PHSs)
		\cite[pp.~116--117]{vdS17} with quadratic, positive definite Hamiltonian function \cite[p.~136]{vdS17}\footnote{Note that I) and II) are equivalent, as every linear passive system with quadratic, positive definite storage function can be written as linear ISO-PHS \cite[pp.~116--117]{vdS17}}. If the dissipation matrix is positive definite, the ISO-PHS is furthermore strictly EIP.
		\item[III)] static, monotonically increasing nonlinearities \cite[p.~11]{Stan07} \cite[p.~24]{Arcak16} \cite[Eq.~(10)]{SimpsonPorco19}
		\begin{align}
			\vec{z}_i = \vec{f}_i(\vec{u}_i)
		\end{align} 
		Strict EIP follows for strict monotonicity.
	\end{enumerate}
\end{lemma}
%
%
\subsubsection{Skew-symmetric interconnections}
Suppose the $N$ subsystems \eqref{eq:prems:subsystem} are coupled via
\begin{equation}
\label{eq:prems:skew_symm_inter}
\vd_i = \sum_{j \in \mathcal{N}_i} \matrix{\phi}_{ij} \z_j=  \sum_{j \in \mathcal{N}^+_i}  \matrix{\phi}_{ij} \z_j- \sum_{j \in \mathcal{N}^-_i}  \matrix{\phi}_{ji}^\Transpose \z_j \hspace{3mm} i=1, \cdots, N,
\end{equation}
where $ \matrix{\phi}_{ij}$ are submatrices, possibly scalars, of appropriate dimension of a static interconnection block matrix $\matrix{\Phi}$. Then, their interconnection structure \eqref{eq:prems:static_intercon} with $\matrix{M}=\matrix{\Phi}=-\matrix{\Phi}^\Transpose$ is called skew symmetric \cite[p.~17]{Arcak16}. Due to the skew-symmetry, the interconnection \eqref{eq:prems:static_intercon} is power-preserving, i.e.\ the power balance is
\begin{equation} \label{eq:prems:inter_power_preserving}
    \z^\Transpose\vd = \z^\Transpose \Phi \z =0.
\end{equation}

\subsubsection{Modular Stability Analysis}
By calling into use the notion of EIP along with skew-symmetry of interactions, one can investigate, without its explicit knowledge, the stability of any feasible equilibrium
\begin{align}\label{eq:prems:unknown_equilibrium}
	\xRef = \left[ \xRef_1, \dots, \xRef_N \right]^\Transpose
	\in \bar{\XX} \subset \Reals^n
\end{align}
of the autonomous, composite system \eqref{eq:prems:comp_system}.
\begin{lemma}[{\cite[Theorem 3.1]{Arcak16}}] \label{lemma:modular_stability}
    Assume the autonomous, composite system \eqref{eq:prems:comp_system} admits an equilibrium \eqref{eq:prems:unknown_equilibrium}. If each subsystem \eqref{eq:prems:subsystem} is EIP with positive definite storage function $S_i(\x_i,\xRef_i) \succ 0$ and their interconnection \eqref{eq:prems:static_intercon} is skew-symmetric, then $\xRef$ is stable with Lyapunov function 
    \begin{align}  \label{eq:prems:composite_lyapunov_func}
	S(\x,\xRef) = \sum_{i=0}^N S_i(\x_i,\xRef_i) \succ 0,\quad
	\dot{S}(\x,\xRef) \leq 0.
    \end{align}
    If all subsystems are striclty EIP with $S_i(\x_i,\xRef_i) \succ 0$ , asymptotic stability follows.
\end{lemma}

\section{ Modeling and Problem Formulation} \label{sec:Modelling}
In this section, we start by setting out the model of an ImG with arbitrary, time-varying topology.
In parts, the individual models already exist in literature, although scattered and with different features and inconsistent notation. Thus, we combine them here to provide a consistent, unified basis for the subsequent main results in the stability analysis in Section~\ref{sec:stability}.
Section~\ref{sec:modelling:img} formally introduces the ImG as weakly connected, bipartite digraph without self-loops. Sections~\ref{sec:modelling:dgu} -- \ref{sec:modelling:line} present the models of the three main subsystems \emph{DGU}, \emph{load}, and \emph{power line}.
In Section \ref{sec:problem}, we then formulate the main problem of this work which is \emph{asymptotic voltage and frequency stability} in ImGs. 
For the modeling, we make the following usual assumption to use models in the \dq frame rotating at $\freqRef$ (cf.\
\cite{Schiffer16,Etemadi12a,Babazadeh13,Riverso15,Tucci17ac,Sadabadi17, Akagi07, Machowski08,Baimel17}):
\begin{assumption} \label{assumption:balanced_three_phase}
Three-phase electrical signals are balanced and thus without zero-sequence.
\end{assumption}
%
%
Furthermore, under normal grid conditions, the following assumption is valid:
\begin{assumption} \label{assumption:voltage_amp}
Any voltage \emph{amplitudes} (node, reference, nominal) are strictly positive, i.e.\
	\begin{equation}\label{eq:assumption:voltage_amp}
	\Vamp(t) = \sqrt{\Vd(t)^2 + \Vq(t)^2} \in \Reals_{>0}, \quad \forall t \geq 0
	\end{equation}
with $\Vdq(t)\in \Reals^2 \backslash \{\zero_2\}$. The reference frequency is strictly positive $\freqRef \in \Reals_{>0}$. All load model parameters are positive, i.e. real numbers greater or equal to zero.
\end{assumption}

\subsection{Islanded AC Microgrid Model} \label{sec:modelling:img}
\begin{figure} 
	\centering
	\scalebox{.56}{






\begin{tikzpicture}
	\newcommand{\dguTextDGU}[1]{DGU ${#1}$}
	\newcommand{\dguTextZIP}[1]{ZIP ${#1}$}
	\newcommand{\dguTextEXP}[1]{EXP ${#1}$}
	\newcommand{\lineText}[2]{Line ${#1}$ \\ \small (#2 \si{\kilo\meter})}
	
	\def\dguXdist{4.5cm}
	\def\dguYdist{3.5cm}
	\def\dguNodes{3,6}
	\def\dguLoadNodesZIP{1,2,4,5}
	\def\loadNodesZIP{7,8,9}
	\def\loadNodesEXP{9,10,11}
	\def\lines{1/7, 7/2, 2/8, 8/3, 2/5, 8/9, 9/4, 3/6, 4/10, 10/5, 5/11, 11/6}
	\def\lineNodesLengths{1/7/12/2.8, 7/2/13/4.4, 2/8/14/0.6, 8/3/15/0.6, 2/5/16/1.3, 8/9/17/0.5, 9/4/18/0.3, 3/6/19/1.5, 4/10/20/0.8, 10/5/21/0.3, 5/11/22/1.7, 11/6/23/0.2}

	\tikzstyle{dgu}    = [draw, rectangle, rounded corners=1.5mm, minimum height = 1.1cm, fill={red!30}]
	\tikzstyle{dguload}= [draw, rectangle split, rounded corners, rectangle split parts=2, rectangle split part fill={red!30, orange!30}]
	\tikzstyle{load}   = [draw, rectangle,rounded corners=1.5mm, minimum height = 1.1cm, fill={orange!30}]
	\tikzstyle{line}   = [draw, rounded rectangle, minimum height = 0.6cm, fill={blueDark!40!white},text width=1.5cm,align=center]
	\tikzstyle{arrowInFilled}  = [-latex, line width=1.0pt]
	\tikzstyle{arrowOutFilled} = [-latex, line width=1.0pt]
	\tikzstyle{arrowInDashed}    = [-latex, dotted, line width=1.2pt]
	\tikzstyle{arrowOutDashed}   = [-latex, dotted, line width=1.2pt]
	
	\coordinate(cN1);
	\path (cN1) +(0,-\dguYdist*3/4) coordinate 			(cN7);
	\path (cN7) +(0,-\dguYdist*3/4) coordinate 			(cN2);
	\path (cN2) +(-\dguXdist,-\dguYdist*3/4) coordinate (cN8);
	\path (cN8) +(0,-1.35*\dguYdist) coordinate 	(cN3);	
	\path (cN8) +(\dguXdist,0) coordinate  			(cN9);
	\path (cN9) +(0,-\dguYdist) coordinate  		(cN4);
	\path (cN4) +(\dguXdist,0) coordinate  			(cN10);
	\path (cN10)+(0,\dguYdist) coordinate  			(cN5);
	\path (cN5) +(\dguXdist,0) coordinate  			(cN11);
	\path (cN11) +(0,-1.35*\dguYdist) coordinate  		(cN6);
	
	\foreach \a/\b in \lines {\coordinate (cLine\a\b) at ($(cN\a)!0.5!(cN\b)$);}
	
	\foreach \a in \dguNodes {\node[dgu](node\a) at(cN\a) {\dguTextDGU{\a}}; }
	
	\foreach \a in \dguLoadNodesZIP {\node[dguload](node\a) at(cN\a) {\dguTextDGU{\a}\nodepart{second}\dguTextZIP{\a}}; }
	
	\foreach \a in \loadNodesZIP {\node[load](node\a) at(cN\a) {\dguTextZIP{\a}}; }
	\foreach \a in \loadNodesEXP {\node[load](node\a) at(cN\a) {\dguTextEXP{\a}}; }
	
	\foreach \a/\b/\n/\l in \lineNodesLengths {\node[line](line\a\b) at(cLine\a\b){\lineText{\n}{\l}};}
	
	\foreach \a/\b in \lines {
		\draw[arrowInFilled](node\a) to (line\a\b);
		\draw[arrowOutFilled] (line\a\b) to (node\b);	}
	
	
	%
	%
\end{tikzpicture}
}
	\caption{Bipartite graph representation of Feeder 1 of the CIGRE medium voltage distribution network benchmark \cite{Strunz14} as ImG comprising ZIP and EXP loads; DGUs are added at nodes $\DD=\{1,2,3,4,5,6 \}$}
	\label{fig:CIGRE_microgrid}
\end{figure}
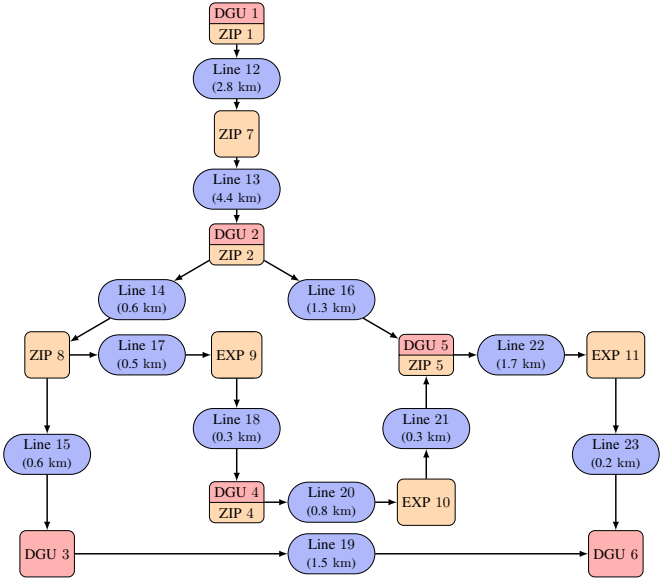

The ImG is modeled as a weakly connected digraph $\GG=(\VV, \EE)$ without self-loops as illustrated in Fig.~\ref{fig:CIGRE_microgrid}. The nodes $\VV$ are partitioned into two sets: $\MM=\{1,\dots,D+L\}$ represents the DGUs and loads, and $\PP= \{D+L+1,\cdots,D+L+P\}$ the power lines. The set $\MM$ is further divided into $\MM=\DD\cup\LL$, where $\DD=\{1,\dots,D\}$ is the set of DGUs, which may each supply a local load, and $\LL= \{D+1,\cdots,D+L\}$ is the set of lone-standing loads. Since DGUs and lone-standing loads are always connected via power lines, all edges in $\EE$ have exactly one node in $\MM$ and another in $\PP$, making $\GG$ a bipartite graph.
The orientation of each edge represents the reference direction of positive line currents. 

\subsubsection{Dynamic model of a DGU} \label{sec:modelling:dgu}
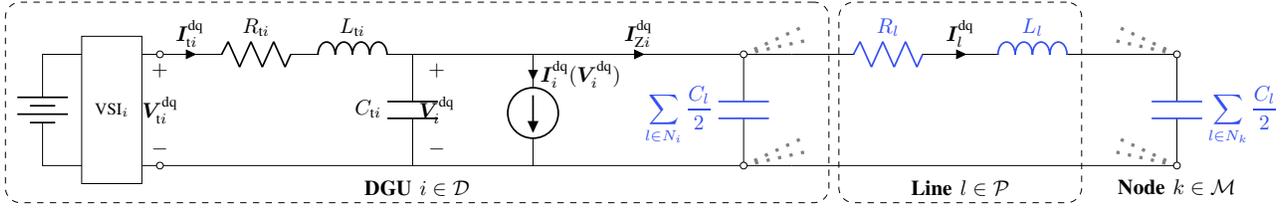
\begin{figure*}
	\centering
	\scalebox{.8}{

\begin{tikzpicture}
	\def\cHeight{1.85cm}	
	\coordinate(dgu_filter_base);
	\draw
		(dgu_filter_base) to [open, o-, v<=${\Vtdq[i]}$] ++(0,\cHeight) coordinate(dgu_filter_base_high)
		to [short, o- ,i=${\Itdq[i]}$] ++(1.0,0)
		to [R, l=${\Rti}$] ++(1.2,0) 
		to [L, l=${\Lti}$] ++(2.0,0) coordinate(dgu_filter_output_high) 
		to [C, l_=${\Cti}$] ++(0,-\cHeight) coordinate(dgu_filter_output_low) 
		to [short] (dgu_filter_base)
		
		(dgu_filter_output_high) to [open] ++(0.4,0)
		to [open, v^>=${\Vdq_i}$] ++(0,-\cHeight)
		
		(dgu_filter_output_high) to  ++(2,0) coordinate(dgu_load_high)
		to [short] ++(0,-0.1cm)
		to [american current source, i>^=${\Idq_i(\Vdq_i)}$] ++(0,{-\cHeight+0.1cm}) coordinate(dgu_load_low)
		to [short] (dgu_filter_output_low)
		
		(dgu_load_high) to [short, i^>=${\Izdq[i]}$,-o] ++(3.5, 0) coordinate(dgu_c_parallel_high)
		to [C, color=\colorLine, \colorLine, l_=$\displaystyle {\color{\colorLine}\sum_{l \in N_i}\dfrac{C_\sLine}{2}}$] ++(0, -\cHeight) coordinate(dgu_c_parallel_low)
		to [short, o-] (dgu_load_low)
		
		(dgu_c_parallel_high) to [short] ++(0.8, 0) coordinate(dgu_pcc_high)
		++(0, -\cHeight) coordinate(dgu_pcc_low)
		to [short] (dgu_c_parallel_low)
		
		(dgu_filter_base) to [short] ++(-1.95,0) coordinate(dgu_source_low)
		to [battery, invert] ++(0,\cHeight) coordinate(dgu_source_high)
		to [short] (dgu_filter_base_high);
	
	\path (dgu_filter_base) +(-0.8,\cHeight/2) coordinate (inverter);
	\node[align=center,draw,fill=white,rectangle,minimum height =\cHeight + 0.6cm,minimum width =1.0cm ]() at(inverter) {\footnotesize VSI$_i$};
	
	

	\coordinate(line_left_low) at (dgu_pcc_low);
	\coordinate(line_left_high) at (dgu_pcc_high);
	\draw	
		
		
		(line_left_high) to [short] ++(1.0,0)
		to [R, color=\colorLine, \colorLine, l=$R_\sLine$] ++(1.2,0)
		to [short, i^>=${\Idq_\sLine}$] ++(1.2, 0)
		to [L, color=\colorLine, \colorLine, l=$L_\sLine$] ++(1.2,0)
		to [short] ++(0.8,0) coordinate(line_Cj_high)
		++(0,-\cHeight) coordinate(line_Cj_low)
		to [short] (line_left_low)
		
		(line_Cj_high) to [short, -o] ++(1.0,0) coordinate(line_right_high)
		to [C, color=\colorLine, \colorLine, l=$\displaystyle {\color{\colorLine}\sum_{l \in N_k}\dfrac{C_\sLine}{2}}$]
		++(0,-\cHeight) coordinate(line_right_low)
		to [short, o-] (line_Cj_low);
		
	\node[below=0.11] at (line_right_low) {\textbf{Node $k \in \MM$}};

	\path (dgu_source_low) +(-0.5,-0.5) coordinate (dgu_box_bottom_left);
	\path (line_left_high) +(0.5,0.75) coordinate (dgu_box_top_right);
	
	\path (line_left_low) +(0.9, -0.5) coordinate (line_box_bottom_left);
	\path (line_Cj_high) +(-0.7, 0.75) coordinate (line_box_top_right);
	
	\begin{scope}[on background layer]
		\node[draw,dashed,rounded corners=0.25cm,fit=(dgu_box_bottom_left) (dgu_box_top_right)] (dgu_box) {};
		\node[above] at (dgu_box.south) {\textbf{DGU $i \in \DD$}};
		
		\node[draw,dashed,rounded corners=0.25cm,fit=(line_box_bottom_left) (line_box_top_right)] (line_box) {};
		\node[above] at (line_box.south) {\textbf{Line $l \in \PP$}};
		
	\draw[loosely dotted, line width=1.55pt, draw=black!50]
		(dgu_c_parallel_low) to ++(12.5:1.1)
		(dgu_c_parallel_high) to ++(12.5:1.1)
		
		(dgu_c_parallel_low) to ++(25:1.1)
		(dgu_c_parallel_high) to ++(25:1.1);
		
	\draw[loosely dotted, line width=1.55pt, draw=black!50]
		(line_right_low) to ++(-12.5:-1.1)
		(line_right_high) to ++(-12.5:-1.1)
		
		(line_right_low) to ++(-25:-1.1)
		(line_right_high) to ++(-25:-1.1);
	\end{scope}
		
\end{tikzpicture}}
	\caption{Circuit diagram of a DGU $i \in \DD$ comprising a DC voltage source, a VSI controlled by a local PnP voltage controller, and a series $RLC$ filter connected to three-phase $\Pi$-lines ({\color{\colorLine} blue}) and possibly an AC load (voltage-dependent current sink); the legs of the lines are considered part of the respective subsystems at the nodes}
	\label{fig:dgu}
\end{figure*}

The circuit diagram of a DGU at any node $i \in \DD$ in the ImG is shown in the left dashed frame of Fig.~\ref{fig:dgu}.
%
The DGU comprises a DC voltage source (generally a renewable energy source or storage device),  a VSI in grid-forming mode, and a series $RLC$ filter \cite{Schiffer16, Olivares14, Guerrero13}. The losses in the VSI and filter are lumped together in $\Rti$. The VSI is modeled by an average model commonly employed for control design \cite{Schiffer16}.
The DC source together with the averaged VSI model is assumed to form an ideal three-phase AC voltage source that supplies sufficient power at all times (cf.~\cite{Schiffer16, Riverso15, Tucci17ac,Sadabadi17,Cucuzzella18}). 

On applying Kirchhoff's current law (KCL) and Kirchhoff's voltage law (KVL) to the components in \dq coordinates \cite{Schiffer16, Baimel17}, we obtain the linear ISO-PHS model (cf.\ \cite[pp.~116--117]{vdS17}) as
\begin{subequations}
	\label{eq:modelling:dgu_phs}
	\begin{align} \label{eq:modelling:dgu_phs_dynamics}
			\xdot_i &=  \left[ \J_i - \R_i \right] \frac{\partial \Hamil_i(\x_i)}{\partial \x_i}+\G_i \vu_i + \K_i \vd_i\\
\label{eq:modelling:dgu_phs_y}
			\y_i &= \G_i^\Transpose \frac{\partial \Hamil_i(\x_i)}{\partial \x_i} \\
 \label{eq:modelling:dgu_phs_z}
			\z_i &= \K_i^\Transpose \frac{\partial \Hamil_i(\x_i)}{\partial \x_i}
	\end{align}
with Hamiltonian
	\begin{equation} \label{eq:modelling:dgu_hamiltonian}
		\Hamil_i(\x_i) = \frac{1}{2} \x_i^\Transpose \, \underbrace{\text{Diag}\left[\frac{1}{\Lti}, \frac{1}{\Lti}, \frac{1}{\Cti}, \frac{1}{\Cti}\right]}_{\Q_i} \x_i  ,
	\end{equation}
	states $\x_i = \left[\Lti \Itd[i], \Lti \Itq[i],  \Cti \Vd_i,  \Cti \Vq_i \right]^\Transpose$, co-states $\frac{\partial \Hamil_i(\x_i)}{\partial \x_i}=\left[ \Itd[i], \Itq[i],\Vd_i,\Vq_i\right]^\Transpose$,
	internal control input to the corresponding VSI $\vu_i= \left[\Vtd[i], \Vtq[i]\right]^\Transpose$,
	uncontrollable interaction (coupling) input 
	\begin{equation}
	\label{eq:inputDGU}
	 \vd_i = \begin{bmatrix}
	 -\Izd[i]\\
	 -\Izq[i]
	 \end{bmatrix}
	 = -\sum_{l \in \mathcal{N}_i^+}\Idq_\sLine + \sum_{l \in \mathcal{N}_i^-}\Idq_\sLine,
	\end{equation}
	 internal control output $\y_i=\left[\Itd[i], \Itq[i]\right]^\Transpose$, uncontrollable interaction (coupling) output $\z_i=\left[\Vd_i, \Vq_i\right]^\Transpose$,
	and interconnection, damping, input, and interaction matrices
	\begin{equation} \label{eq:modelling:dgu_JR}
		\begin{alignedat}{3}
			\J_{i} =& \begin{bmatrix}
				0 & \freqRef \Lti & -1 & 0 \\
				-\freqRef \Lti & 0 & 0 & -1 \\
				1 & 0 & 0 & \freqRef \Cti \\
				0 & 1 & -\freqRef \Cti & 0
			\end{bmatrix} \, , \\
			\R_{i} =& \; \text{Diag}\left[\, \Rti, \, \Rti, \, 0, \, 0 \,\right] \, ,\\
			\G_i =& \begin{bmatrix}
				1 & 0 \\
				0 & 1 \\
				0 & 0 \\
				0 & 0
			\end{bmatrix},
			\K_i= \begin{bmatrix}
				0 & 0 \\
				0 & 0 \\
				1 & 0 \\
				0 & 1
			\end{bmatrix}.
		\end{alignedat}
	\end{equation}
\end{subequations}
Note that the pair $(\vd_i,\z_i)$ forms the uncontrollable interaction (coupling) port accounting for the coupling with the remaining ImG. 
\begin{remark} \label{remark:load_at_DGU}
In case DGU $i$ supplies a local, static AC load modeled as voltage-dependent current sink, one can obtain the new dynamics by subtracting $\K_i\RR_i(\bm{x}_i)$ from \eqref{eq:modelling:dgu_phs_dynamics}, where $\RR_i(\x_i)=\Idq_i(\Vdq_i)$ is for example modeled as \eqref{eq:modelling:nonlinear_R_zip_exp}.
\end{remark}

\subsubsection{Nonlinear Static AC Loads} \label{sec:modelling:load}
\begin{figure}
	\centering
	\scalebox{0.9}{\begin{tikzpicture}
	\def\cHeight{2.5cm}		
	
	\coordinate(port_left_low);
	\draw
		(port_left_low) to [open] ++(0,\cHeight) coordinate(port_left_high)
		to [short, o-] ++(0.7, 0) coordinate(line_top_left)
		to [R, color=\colorLine,\colorLine, l=${R_\sLine}$] ++(1.2,0)
		to [short, i^>=${\Idq_\sLine}$] ++(1.0, 0)
		to [L, color=\colorLine,\colorLine, l=${L_\sLine}$] ++(1.2,0) coordinate(line_top_right)
		to [short] ++(0.7,0) coordinate(port_right_high)
		to [open] ++(0,-\cHeight) coordinate(port_right_low)
		to [short, -o] (port_left_low)
		
		(line_top_right) to [open] ++(0,-\cHeight) coordinate(line_bottom_right)

		(port_right_high) to [short,o-,i^<=${\Izdq[j]}$] ++(2.6,0) coordinate(line_C_high)
		to [C, align=center,  l_=${C_j=}$\\ $\displaystyle {\color{\colorLine}\sum_{l \in N_j}\dfrac{C_\sLine}{2}}$ ] ++(0,-\cHeight) coordinate(line_C_low)
		to [short,-o] (port_right_low)
		
		(line_top_left) to [open] ++(-0.5,0)
		to [open, v^>=${\Vdq_k}$] ++(0, -\cHeight)
		(port_right_high) to [open] ++(0.2,0)
		to [open, v_>=${\Vdq_j}$] ++(0, -\cHeight);
	
	\draw
		(line_C_high) to [short, i>^=${\Idq_j({\Vdq_j})}$] ++(1.25cm,0) coordinate(load_top)
		to [short] ++(0,-0.1cm)
		to [american current source] ++(0,{-\cHeight+0.1cm}) coordinate(load_bottom)
		to [short] (line_C_low);
		

	\path (line_top_left) +(-0.35,0.75) coordinate (line_box_top_left);
	\path (line_bottom_right) +(0.9,-0.45) coordinate (line_box_bottom_right);
	\path (load_bottom) +(0.65,-0.45) coordinate (load_box_bottom_right);
	\path (port_right_high) +(0.75,0.75) coordinate (load_box_top_left);
	
	\begin{scope}[on background layer]
		\node[draw,dashed,rounded corners=0.25cm,fit=(line_box_bottom_right) (line_box_top_left)] (line_box) {};
		\node[above] at (line_box.south) {\textbf{Line $l \in \PP$}};
		\node[draw,dashed,rounded corners=0.25cm,fit=(load_box_bottom_right) (load_box_top_left)] (load_box) {};
		\node[above] at (load_box.south) {\textbf{Load $j \in \LL$}};
		
	\draw[loosely dotted, line width=1.55pt, draw=black!50]
		(port_right_low) to ++(14.5:-1.1)
		(port_right_high) to ++(14.5:-1.1)
		
		(port_right_low) to ++(25:-1.1)
		(port_right_high) to ++(25:-1.1);
	\end{scope}
		
\end{tikzpicture}}
	\caption{Circuit diagram of a, lone-standing, nonlinear static AC load $j\in \LL$ connected to three-phase $\Pi$-lines ({\color{\colorLine}blue}); $C_j$ is the sum of the parallel capacitances ${\color{\colorLine}\frac{C_\sLine}{2}}$ of the connected lines}
	\label{fig:load}
\end{figure}
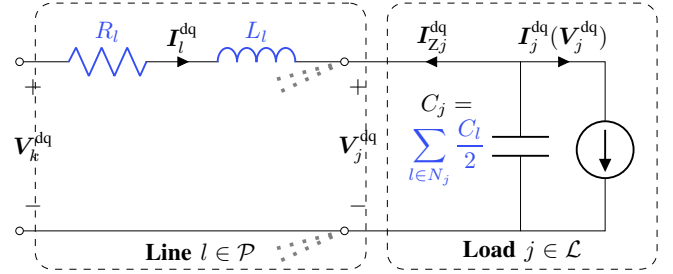

The circuit diagram of a lone-standing AC load at any node $j\in\LL$ along with connecting power lines is shown in Fig.~\ref{fig:load}. The load is modeled as voltage-dependent current sink representing a nonlinear static AC load. By applying KCL to the components in \dq coordinates \cite{Schiffer16, Baimel17}, we obtain the ISO-PHS model with nonlinear resistive structure (cf.\ \cite[p.~114]{vdS17})
\begin{subequations} \label{eq:modelling:loadphs}
	\begin{align}
		\xdot_\sLoad&=
		\JL \frac{\partial \Hamil_\sLoad(\xL)}{\partial \xL}-\RL(\xL)+
		\KL \vd_\sLoad \\
		\z_j&=\KL^\Transpose \frac{\partial \Hamil_\sLoad(\xL)}{\partial \xL} \label{eq:modelling:loadphs_z}\\
		\label{eq:modelling:loadphs:hamil}
		\Hamil_\sLoad(\xL)&=\frac{1}{2} \xL^\Transpose \text{Diag}\left[\frac{1}{C_\sLoad}, \frac{1}{C_\sLoad} \right]  \xL
	\end{align}
	with states $\xL=\left[C_\sLoad \VLd, C_\sLoad \VLd \right]^\Transpose$, co-states $\frac{\partial \Hamil_\sLoad(\xL)}{\partial \xL} = \left[\VLd, \VLq \right]^\Transpose$ uncontrollable interaction (coupling) input 
	\begin{equation}
	    \label{eq:inputLoad}
	    \vd_\sLoad= \begin{bmatrix}
	    -\Izd[j]\\
	     -\Izq[j]
	    \end{bmatrix}
	    = -\sum_{l \in \mathcal{N}_j^+}\Idq_\sLine + \sum_{l \in \mathcal{N}_j^-}\Idq_\sLine,
	\end{equation}
	interaction (coupling) output $\z_j=\left[\VLd, \VLq \right]^\Transpose$, nonlinear resistive structure
	\begin{equation}\label{eq:modelling:nonlinear_R}
		\RL(\xL)=\ILdq(\VLdq),
	\end{equation}
	and interconnection and interaction matrices
	\begin{equation}\label{eq:modelling:phs_matrices}
	\JL=\begin{bmatrix}
	0				&	\freqRef C_\sLoad\\
	-\freqRef C_\sLoad	&	0
	\end{bmatrix}, \quad
	\KL=\begin{bmatrix}
	1	&	0\\
	0	&	1
	\end{bmatrix}.
	\end{equation}
\end{subequations}
The nonlinear resistive structure \eqref{eq:modelling:nonlinear_R} is specified by the nonlinear, static \dq load current-voltage relation of the considered AC load models. Prevalent such models are \emph{constant impedance\footnote{Note that constant impedances (Z) are actually expressed as admittances (Y)}, constant current, constant power} (ZIP) loads
\begin{subequations} \label{eq:modelling:zip}
	\begin{align}
		\lAct(\Vamp[\sLoad]) &= \lYp[\sLoad] \Vamp[\sLoad]^2 + \lIp[\sLoad] \Vamp[\sLoad] + \lPp[\sLoad] \, , \label{eq:modelling:zip:act} \\
		\lReac(\Vamp[\sLoad]) &= \lYq[\sLoad] \Vamp[\sLoad]^2 + \lIq[\sLoad] \Vamp[\sLoad] + \lPq[\sLoad] \, , \label{eq:modelling:zip:reac}
	\end{align}
\end{subequations}
and \emph{exponential} (EXP) loads
\begin{subequations} \label{eq:modelling:exp}
	\begin{align}
		\lAct(\Vamp[\sLoad]) &= \lActNom[\sLoad] \left(\frac{\Vamp[\sLoad]}{\VampNominal}\right)^{\lNp[\sLoad]} \, , \\
		\lReac(\Vamp[\sLoad]) &= \lReacNom[\sLoad] \left(\frac{\Vamp[\sLoad]}{\VampNominal}\right)^{\lNq[\sLoad]} \, ,
	\end{align}
\end{subequations}
where $\VampNominal$ is the nominal phase-to-phase RMS value (e.g.\ \SI{400}{\volt}), $\Vamp[\sLoad]$ is the amplitude of the instantaneous, complex \dq voltage vector (see \eqref{eq:assumption:voltage_amp}) equaling the phase-to-phase RMS voltage in steady-state, $\{\lYp[\sLoad],  \lIp[\sLoad], \lPp[\sLoad]\}$ and $\{ \lYq[\sLoad],  \lIq[\sLoad], \lPq[\sLoad]\}$ are the respective active and reactive ZIP parameters, and $\lActNom[\sLoad]$ and $\lReacNom[\sLoad]$ are the nominal active and reactive powers with respective voltage indexes $\lNp[\sLoad]$ and $\lNq[\sLoad]$ \cite[pp.~111--112]{Machowski08}\cite[pp.~33--34]{Farrokhabadi18techrep}. 
However, the active and reactive power representations \eqref{eq:modelling:zip} and \eqref{eq:modelling:exp} commonly found in pertinent literature do not coincide with the first-principles (currents, voltages) modeling required in \eqref{eq:modelling:nonlinear_R}. Thus, in \cite[Lemma~1]{Strehle20ifac} it was shown how \eqref{eq:modelling:zip} and \eqref{eq:modelling:exp} can be reformulated to obtain a differentiable load current mapping $\ILdq(\Vdq_j): \Reals^2 \backslash \{\zero_2\} \to \Reals^2$
\begin{equation}\label{eq:modelling:nonlinear_R_zip_exp}
	\ILdq(\VLdq)=
	\begin{dcases}
	\footnotesize
	\begin{bmatrix}
	\Bigg. \dfrac{\lPp[\sLoad] \VLd + \lPq[\sLoad] \VLq}{\Vamp[\sLoad]^2} + \dfrac{\lIp[\sLoad] \VLd + \lIq[\sLoad] \VLq}{\Vamp[\sLoad]} \\[-4pt]
	\Big. + \lYp[\sLoad] \VLd + \lYq[\sLoad] \VLq \\
	\hdashline
	\Bigg. \dfrac{\lPp[\sLoad] \VLq - \lPq[\sLoad] \VLd}{\Vamp[\sLoad]^2} + \dfrac{\lIp[\sLoad] \VLq - \lIq[\sLoad] \VLd}{\Vamp[\sLoad]} \\[-4pt]
	\Big. + \lYp[\sLoad] \VLq - \lYq[\sLoad] \VLd
	\end{bmatrix}& \text{ZIP}\\
	\begin{bmatrix}
		\Bigg. \dfrac{\lActNom[\sLoad] \Vamp[\sLoad]^{\lNp[\sLoad] - 2}}{\VampNominal^{\lNp[\sLoad]}} \VLd + \lReacNom[\sLoad] \dfrac{\Vamp[\sLoad]^{\lNq[\sLoad] - 2}}{\VampNominal^{\lNq[\sLoad]}} \VLq \\
		\hdashline
		\Bigg. \dfrac{\lActNom[\sLoad] \Vamp[\sLoad]^{\lNp[\sLoad] - 2}}{\VampNominal^{\lNp[\sLoad]}} \VLq - \lReacNom[\sLoad] \dfrac{\Vamp[\sLoad]^{\lNq[\sLoad] - 2}}{\VampNominal^{\lNq[\sLoad]}} \VLd
	\end{bmatrix} & \text{EXP}
	\end{dcases}
\end{equation}

\subsubsection{Dynamic model of a power line}  \label{sec:modelling:line}
Any power line $l \in \PP$ from node $a\in \MM$ to $b\in \MM$ is represented by the $\Pi$-equivalent model of a transmission line \cite[p.~201--207]{Kundur94} with $C_\sLine,R_\sLine,L_\sLine>0$ as illustrated in Fig.~\ref{fig:dgu} and \ref{fig:load}. The C legs on both sides of the line are considered to be lumped into the capacitances $C_{\sInput a}, C_{\sInput b}, C_a, C_b$ of the respective node subsystem $a,b \in \MM$ (see Fig.~\ref{fig:dgu} and \ref{fig:load}). This avoids dependent storages (i.e.\ parallel capacitances in this case) and thus dependent states, which lead to implicit models that can only be made explicit in specific cases \cite[p.~107,110]{Borutzky10}\cite[p.~53]{vdS14}\cite{Pfeifer2020dependent}.
By applying KVL in \dq coordinates \cite{Schiffer16, Baimel17} on the resulting RL model of the $l^{\text{th}}$ power line, we obtain the linear ISO-PHS 
\begin{subequations} \label{eq:modelling:line_phs}
	\begin{align} \label{eq:modelling:line_phs_dynamics}
	\xdot_l &=  \left[ \J_l - \R_l \right] \frac{\partial \Hamil_l(\x_l)}{\partial \x_l} + \K_l \vd_l\\
	\z_l &= \K_l^\Transpose \frac{\partial \Hamil_l(\x_l)}{\partial \x_l} \label{eq:modelling:line_phs_z}
	\end{align}
	with Hamiltonian
	\begin{equation} \label{eq:modelling:line_hamiltonian}
		\Hamil_\sLine( \x_{\sLine}) = \frac{1}{2} \x_{\sLine}^{\Transpose} \, \text{Diag}\left[\frac{1}{ L_\sLine}, \frac{1}{L_\sLine} \right] \x_{\sLine} \, ,
	\end{equation}
	states $\x_{\sLine} = \left[L_\sLine \Ilined, \, L_\sLine \Ilineq\right]^{\Transpose}$, co-states $\frac{\partial \Hamil_l(\x_l)}{\partial \x_l}= \left[ \Ilined, \, \Ilineq\right]^{\Transpose}$ uncontrollable interaction (coupling) input
\begin{equation} \label{eq:inputLines}
    \vd_{\sLine} = 
    \begin{bmatrix}
    \Vdq_{a}\\
    \Vdq_{b}
    \end{bmatrix}=
    \sum_{k \in \mathcal{N}_l^+}
    \begin{bmatrix}
    \Zero_2\\
    \One_2
    \end{bmatrix} \Vdq_k +
    \sum_{k \in \mathcal{N}_l^-}
    \begin{bmatrix}
    \One_2\\
    \Zero_2
    \end{bmatrix} \Vdq_k,
\end{equation}
interaction (coupling) output $\z_{\sLine} = \left[\Ilinedq, -\Ilinedq\right]^{\Transpose}$,	interconnection, damping, and interaction matrices
	\begin{equation} \label{eq:modelling:line_JR}
	\begin{split}
			\J_{\sLine} &= \begin{bmatrix}
				0 & \freqRef L_\sLine \\
				-\freqRef L_\sLine & 0 
			\end{bmatrix}, \;
			\R_{\sLine} = \begin{bmatrix} 
				R_\sLine & 0 \\
				0 & R_\sLine 
			\end{bmatrix},\\
			\K_l&=\begin{bmatrix}
			\One_{2 \times 2} & -\One_{2 \times 2}
			\end{bmatrix}.
			\end{split}
	\end{equation}
\end{subequations}
 Note that \eqref{eq:modelling:line_phs} has no controllable ports. 
\subsection{Problem Formulation} \label{sec:problem}
In order to formulate the main control objective of \emph{asymptotic voltage and frequency stability}, we recall that a variable $\x_{\sDQ}(t) = \left[ \x_\sD(t), \x_\sQ(t) \right]^\Transpose$ in the \dq frame rotating at frequency $\freqRef$ can be considered a Cartesian representation of a complex vector with polar representation \cite{Schiffer16}
\begin{equation} \label{eq:stability:phasor_def}
    A(t) \angle \theta(t) \text{ with }\begin{cases}
      A(t)=\sqrt{x_\sD^2(t)+x_\sQ^2(t)}  \\
        \theta(t)= \mathrm{arctan}(\frac{x_\sQ(t)}{x_\sD(t)})
    \end{cases} ,
\end{equation}
often referred to as phasor \cite[p.~60]{Machowski08}. 
From \eqref{eq:stability:phasor_def}, it is evident that an asymptotically stable equilibrium $\xRef_{\sDQ}=const.$ implies
\begin{align}
    \lim\limits_{t \to \infty} A(t)&=const.\\
    \lim\limits_{t \to \infty} \theta(t)&=const. \label{eq:stability:theta_const}
\end{align}
Thus, if the various nodal \dq frames are sufficiently synchronized to the desired $\freqRef$, asymptotic stability of the \dq node voltage equilibria $\bar{\vec{V}}^{\sDQ*}_k, k \in \MM$ also guarantees an ImG-wide asymptotically stable frequency equilibrium $\freqRef$. Technologies for achieving such synchronization via infrequent communication are available (see \cite{Etemadi12a}, \cite[Assumption~2]{Cucuzzella18} or the review in \cite{Tucci17ac, Tucci18TechRep}). Thus, we establish the following assumption:
\begin{assumption} \label{assumption:dq_synchrony}
	The \dq reference frames at all nodes $k \in \MM$ are synchronized. 
\end{assumption}
The main control problem addressed in this work now reads as follows:
\begin{problem} \label{problem}
Consider an ImG as modeled in Section~\ref{sec:modelling:img} and which fulfills Assumptions~\ref{assumption:balanced_three_phase}, \ref{assumption:voltage_amp}, and \ref{assumption:dq_synchrony}. When is the set of \dq node voltage equilibria $\bar{\vec{V}}^{\sDQ*}_k, k \in \MM=\DD \cup \LL$ containing known equilibria $\VdqRef_i, i \in \DD$ and unknown equilibria $\bar{\vec{V}}^\sDQ_j, j \in \LL$ asymptotically stable?

\end{problem}
%
%
\begin{remark}
Note that the \dq reference frames at load nodes $j \in \LL$ are only used as means to facilitate the analysis of the balanced, three-phase AC signals. There are no actual controller clocks to be synchronized. 
\end{remark}
\section{Control Design} \label{sec:control}
Given Lemma~\ref{lemma:modular_stability}, Problem~\ref{problem} can be addressed in a modular manner by analyzing the EIP of the subsystems \emph{DGU}, \emph{load}, and \emph{power line}, and the skew-symmetry of their interconnections.
%
Under the roof of this very general EIP framework, a variety of control designs for the DGU VSIs are possible which render the closed-loop system EIP with respect to the desired equilibrium
\begin{equation} \label{eq:control:DGU_equilibrium_desired}
    \xRef^*_i =\left[\Lti \IdRef[i], \Lti \IqRef[i], \Cti \VdRef_i, \Cti \VqRef_i \right]^\Transpose \in \bar{\XX}_i, \; i \in \DD.
\end{equation}
\begin{assumption} \label{assumption:higher_level_control}
The voltage references $\VdqRef_i\in\Reals^2\backslash \{\zero_2\}$ are known and specified by a higher-level control ensuring sensible ImG operation and accounting for VSI limitations. The steady-state currents $\IdqRef[i] \in \Reals^2, i \in \DD$, on the other hand, are not known a priori and follow as a consequence of the load demand and network exchange currents (cf.\ \eqref{eq:modelling:dgu_phs}).
\end{assumption}
While there is a rich body of literature in this field for islanded DC microgrids, there are, to the best of the authors' knowledge, only two previous works in the field of AC ImGs.
An IDA-PBC approach based on PHSs whose physical insights allow for a constructive control design and immediate system analysis \cite{Strehle19}; EIP of the closed-loop can, for example, easily be inferred as illustrated in Section~\ref{sec:stability:eip_dgu}.
A Lyapunov approach with an a priori specified PI control structure \cite{Nahata19ecc}; the appropriate tuning of the PI control gains ensures EIP of the closed-loop as illustrated in Section~\ref{sec:stability:eip_dgu_pulkit}.
In this work, we thus use these controllers as examples to demonstrate the basic, unifying features of the EIP framework, which is, however, not restricted to any specific controller type (see also Remark~\ref{remark:controller_guidelines}). For a a consistent and self-contained basis in the subsequent stability analysis in Section~\ref{sec:stability}, we shortly recall the main results from \cite{Strehle19, Nahata19ecc} in Sections \ref{control:felix} and \ref{control:pulkit}. In Section \ref{control:design:integralAction}, we furthermore extend our previous results from \cite{Strehle19} by integral action on the non-passive output. This allows for additional damping while preserving zero steady-state voltage errors and robustifies our design against parameter uncertainties.

\subsection{Port-Hamiltonian Control Design} \label{control:felix}
The control design combines non-parameterized IDA-PBC \cite{Ortega04} with the systematic approach in \cite{Kotyczka13} that splits the system to be controlled into \emph{actuated ($\alpha$)} and \emph{unactuated ($\nu$)} parts. Note that as usual, the design is initially carried out for the unconnected DGU system, i.e.\ the interaction port of the DGU ISO-PHS \eqref{eq:modelling:dgu_phs} is set to zero $\vd_i=\Izdq[i] = \zero_2$.
\begin{remark} \label{remark:drop_subscript_i}
	For clarity of presentation in the subsequent design, the subscript $i$ is dropped from all variables and parameters in this section, i.e.\ $\Vdq := \Vdq_i$ and $\Rt := \Rti$ etc. and the results hold for $i\in\DD_\sPHS$
\end{remark}
\begin{proposition}[\cite{Strehle19}]
Consider the DGU ISO-PHS \eqref{eq:modelling:dgu_phs} with $\vd=\Izdq[] = \zero_2$. Then, the control law
\begin{equation} \label{eq:control:result:u_simple}
	\underset{(2\times 1)}{\vec{\beta}(\x)} \, {=} \left[ \begin{array}{@{\,}c@{\,}}
		\frac{\alpha_{11}}{\nu_{11}} (\Id+\freqRef \Ct \VqRef)\\ 
		- \nu_{11} (\Vd - \VdRef) + \Rt \Id - \freqRef \Lt \Iq + \Vd  \Big.	\\
		\hdashline
		\Big. \frac{\alpha_{22}}{\nu_{22}} (\Iq-\freqRef \Ct \VdRef)  \\ 
		- \nu_{22} (\Vq - \VqRef) + \Rt \Iq + \freqRef \Lt \Id + \Vq
	\end{array} \right],
\end{equation}
with control parameters $ \alpha_{11}, \alpha_{22} <0$ and $\nu_{11}, \nu_{22} > 0 $, references $ \VdRef, \VqRef, \freqRef $, and measurements of the states (or rather their corresponding currents and voltages $ \Vd, \Vq, \Id, \Iq $) yields the closed-loop DGU PHS dynamics
\begin{subequations}
	\label{eq:control:closed_loop_phs}
	\begin{align} 
	\xdot &=  \left[ \J_\sDesired - \R_\sDesired  \right] \frac{\partial \Hd(\x)}{\partial \x}\\
	\Hd(\x) &= \frac{1}{2}  \left(\x - \xRef \right)^\Transpose \Q_\sDesired
	\left(\x - \xRef \right)
	\end{align}
with $\x = \left[x_{\sActuated 1}, x_{\sActuated 2}, x_{\sUnactuated 1}, x_{\sUnactuated 2} \right]^\Transpose=\left[\Lt \Id, \Lt \Iq,  \Ct \Vd,  \Ct \Vq\right]^\Transpose$, $\frac{\partial \Hd(\x)}{\partial \x}=\left[ \frac{\Id + \freqRef \Ct \VqRef}{\nu_{11}},  \frac{\Iq + \freqRef \Ct \VdRef}{\nu_{22}},\Vd-\VdRef,\Vq-\VqRef\right]^\Transpose$, and interconnection, damping, and energy matrices
\begin{align}
%
    \J_\sDesired&=
    \begin{bmatrix}
    0&   0           &   -\nu_{11}       &   0\\
    0           &  0 &       0           &   -\nu_{22}\\
    \nu_{11}    &   0           &       0           &   \freqRef \Ct\\
    0           &   \nu_{22}    &   -\freqRef \Ct   &   0
    \end{bmatrix}\\
    \R_\sDesired&= \text{Diag} \left[ -\alpha_{11}, -\alpha_{22}, 0, 0 \right]\\
    \Q_\sDesired&=\text{Diag}\left[\frac{1}{\Lt \nu_{11}}, \frac{1}{\Lt \nu_{22}}, \frac{1}{\Ct}, \frac{1}{\Ct }\right]
\end{align}
\end{subequations}
%
The desired DGU equilibrium \eqref{eq:control:DGU_equilibrium_desired} is established as $\xRef_i=\text{arg min}_{\x} \Hd(\x)$
\end{proposition}

\subsubsection{Integral Action} \label{control:design:integralAction}
From \eqref{eq:control:result:u_simple}, we see that the control law depends on filter parameters ($\Rt, \Lt$) as well as measurements ($\Idq. \Vdq$). Naturally, model knowledge and measurements are never fully accurate. Furthermore, the variable damping assignment of $\alpha_{11}, \alpha_{22} < 0$ introduces steady state voltage errors under non-vanishing interaction currents $\Izdq \neq \zero_2$. This fact can be understood by recalling that the control design leading to \eqref{eq:control:result:u_simple} is carried out for the undisturbed DGU model \eqref{eq:modelling:dgu_phs}. Thus, the unactuated state reference (cf.\ (30) in \cite{Strehle19})
\begin{equation} \label{eq:control:design:current_equilibrium}
	\x^*_\sActuated=
	\begin{bmatrix}
		\Lt I^{\sD *}_{\sInput} \\
		\Lt I^{\sQ *}_{\sInput}
	\end{bmatrix} \coloneqq
	\begin{bmatrix}
	    \Lt \left(-\freqRef \Ct \VqRef \right)\\
        \Lt \left(\freqRef \Ct \VdRef \right)
	\end{bmatrix} \, ,
\end{equation}
used in the control design does not accurately reflect the equilibrium current $\IdqRef$ in the presence of unknown interactions functioning as disturbances (cf.\ the bottom two rows of \eqref{eq:modelling:dgu_phs} when $\Izdq[i] \neq \zero_2$).
A common strategy to address both issues is the addition of a superimposed \emph{integral action} (IA). For this purpose, the control \eqref{eq:control:result:u_simple} is extended to (cf.\ \cite[p.~445]{Ortega04})
\begin{equation} \label{eq:control:ia:general}
    \vu(\x)=\vec{\beta}(\x)+ \vu_\sIA(\x)
\end{equation}
yielding the closed-loop DGU PHS dynamics (cf.\ \eqref{eq:modelling:dgu_phs}, \eqref{eq:control:closed_loop_phs}) 
\begin{subequations}
\label{eq:control:ia:closed_loop_phs}
\begin{align} 
	\xdot &=  \left[ \J_\sDesired - \R_\sDesired  \right] \frac{\partial \Hd(\x)}{\partial \x}+\G \vu_\sIA(\x)+ \K \vd\\
	\z &= \K^\Transpose \frac{\partial \Hd(\x)}{\partial \x}\\
	\Hd(\x) &= \frac{1}{2}  \left(\x - \xRef^* \right)^\Transpose \Q_\sDesired
	\left(\x - \xRef^* \right)
\end{align}
\end{subequations}
%
%
In the context of PHSs, it is furthermore desirable that the IA preserves the PHS structure and thus the passivity-related stability properties of the equilibrium. However, since the interaction $\vd=-\Izdq$ acts on the unactuated part of the system dynamics $\dot{\x}_\sUnactuated=\Ct \Vdqdot$ (cf.\ \eqref{eq:control:ia:closed_loop_phs}), standard integral feedback of the passive output (see \cite{Ortega04}) is not applicable. Instead, IA via state transformation as in \cite{Donaire09} is employed. In line with practice, this IA achieves asymptotic rejection of unknown interactions, i.e.\ disturbances, assuming they are piecewise constant such that the integrators can converge \cite{Donaire09}.
\begin{proposition}
Consider the closed-loop DGU dynamics \eqref{eq:control:ia:closed_loop_phs} obtained with static state feedback  \eqref{eq:control:result:u_simple}. 
Then, the IA
\begin{subequations}\label{eq:control:ia:u_IA}
    \begin{align}
\hspace{-0.5em}	\begin{bmatrix}
		u_{\sIA,1} \\ u_{\sIA,2}
	\end{bmatrix} \!\!&=\!\!
	\begin{bmatrix}
		\frac{\alpha_{11} k_{\sI 1}}{\Ct} \int (x_{\sUnactuated 1}- x_{\sUnactuated 1}^*) \text{d} t + \frac{k_{\sI 1} \Lt}{\Ct} (x_{\sUnactuated 1}^* - x_{\sUnactuated 1})\\
		\frac{\alpha_{22} k_{\sI 2}}{\Ct} \int (x_{\sUnactuated 2}- x_{\sUnactuated 2}^*) \text{d} t + \frac{k_{\sI 2} \Lt}{\Ct} (x_{\sUnactuated 2}^* - x_{\sUnactuated 2})
	\end{bmatrix} \\[+1.1em] \!\!&=\! \!
	\begin{bmatrix}
		\alpha_{11} k_{\sI 1} \int (\Vd- \VdRef) \text{d} t + k_{\sI 1} \Lt (\VdRef - \Vd)\\
		\alpha_{22} k_{\sI 2} \int (\Vq- \VqRef) \text{d} t + k_{\sI 2} \Lt (\VqRef - \Vq)
	\end{bmatrix}
\end{align}
\end{subequations}
asymptotically rejects unknown, piecewise constant interactions, i.e.\ disturbances, $\bar{\vd}=-\bar{\vec{I}}^\sDQ_\sDisturbance$ and establishes the new equilibrium
\begin{equation} \label{eq:control:ia:equilibrium}
    \bar{\x}^* = 
    \begin{bmatrix}
    \bar{\x}_\sActuated^*\\
    \x_\sUnactuated^*
    \end{bmatrix} = 
    \begin{bmatrix}
    \Lt \left(-\freqRef \Ct \VqRef + \bar{I}^\sD_\sDisturbance \right)\\
    \Lt \left(\freqRef \Ct \VdRef + \bar{I}^\sQ_\sDisturbance\right)\\
    \Ct \VdRef\\
    \Ct \VqRef
    \end{bmatrix}
\end{equation}
\end{proposition}
\IEEEproof
Initially, we extend \eqref{eq:control:ia:closed_loop_phs} by two additional integrator states
\begin{equation} \label{eq:control:ia:integrator_states}
    \Zstates_\sExtended = \begin{bmatrix}
    k_{\sI 1} \int \frac{\partial \Hamil_{\sDesired \text{z}}(\z)}{\partial z_{\sUnactuated 1}} \text{d}t\\
     k_{\sI 2} \int \frac{\partial \Hamil_{\sDesired \text{z}}(\z)}{\partial z_{\sUnactuated 2}} \text{d}t
    \end{bmatrix} = \begin{bmatrix}
    \frac{k_{\sI 1}}{\Ct} \int \left(\Zstate_{\sUnactuated 1} - \Zstate_{\sUnactuated 1}^* \right) \text{d}t\\
     \frac{k_{\sI 2}}{\Ct} \int \left(\Zstate_{\sUnactuated 2} - \Zstate_{\sUnactuated 2}^* \right) \text{d}t
    \end{bmatrix}
\end{equation}
and rewrite it in new $\Zstate$-coordinates as
\begin{subequations} \label{eq:control:ia:closed_loop_phs_z_coord}
    \begin{align} \label{eq:control:ia:closed_loop_phs_z_coord:dynamics}
    \dot{\Zstates} &=
    \left[ \J_{\sDesired\sZ}-\R_{\sDesired\sZ} \right] \frac{\partial \Hamil_{\sDesired \sZ}(\Zstates)}{\partial \Zstates}+\begin{bmatrix}
    \K\\
    \Zero_{2\times2}
    \end{bmatrix} \vd\\
    \z-\z^* &= \begin{bmatrix}
    \K\\
    \Zero_{2\times2}
    \end{bmatrix}^\Transpose \frac{\partial \Hamil_{\sDesired \sZ}(\Zstates)}{\partial \Zstates}\\
    \label{eq:control:ia:closed_loop_phs_z_coord:Hamiltonian}
     \Hamil_{\sDesired \sZ}(\Zstates) &= \frac{1}{2} \begin{bmatrix}
     \Zstates_\sActuated - \bar{\Zstates}_\sActuated\\
     \Zstates_\sUnactuated - \Zstates_\sUnactuated^*\\
     \Zstates_\sExtended
     \end{bmatrix}^\Transpose
     \Q_{\sDesired \sZ}
     \begin{bmatrix}
     \Zstates_\sActuated - \bar{\Zstates}_\sActuated\\
     \Zstates_\sUnactuated - \Zstates_\sUnactuated^*\\
     \Zstates_\sExtended
     \end{bmatrix}
\end{align}
with $\Zstates=\left[ \Zstates_\sActuated, \Zstates_\sUnactuated, \Zstates_\sExtended \right]^\Transpose$, $\frac{\partial \Hamil_{\sDesired \sZ}(\Zstates)}{\partial \Zstates}=\left[ \frac{\Zstate_{\sActuated 1} - \bar{\Zstate}_{\sActuated 1}}{\Lt \nu_{11}}, \frac{\Zstate_{\sActuated 2} - \bar{\Zstate}_{\sActuated 2}}{\Lt \nu_{22}}, \frac{\Zstate_{\sUnactuated 1} - \Zstate^*_{\sUnactuated 1}}{\Ct}, \frac{\Zstate_{\sUnactuated 2} - \Zstate^*_{\sUnactuated 2}}{\Ct}, \frac{\Zstate_{\sExtended1}}{k_{\sI 1}}, \frac{\Zstate_{\sExtended 2}}{k_{\sI 2}}  \right]^\Transpose$,  the uncontrollable interaction (coupling) input and output $\vd=-\vec{I}^\sDQ_\sDisturbance$, $\z-\z^*=\frac{\Zstates_\sUnactuated - \Zstates_\sUnactuated^*}{\Ct}=\Vdq-\VdqRef$, and
\begin{align}
    \J_{\sDesired\sZ}&=\left[\begin{array}{cc|c}
        &   &   \Zero_{2\times2}\\
    \multicolumn{2}{c|}{\smash{\raisebox{.5\normalbaselineskip}{$\J_\sDesired$}}} &  \text{Diag}\left[-k_{\sI 1}, -k_{\sI 2} \right]\\
    \hline\\[-\normalbaselineskip]
    \Zero_{2\times2}   & \text{Diag}\left[k_{\sI 1}, k_{\sI 2} \right]        &   \Zero_{2\times2}         
    \end{array} \right],\\
    \label{eq:control:ia:Rcz}
      \R_{\sDesired\sZ}&=\text{Diag} \left[-\alpha_{11}, -\alpha_{22}, 0, 0, 0, 0 \right],\\
      \label{eq:control:ia:Qcz}
      \Q_{\sDesired \sZ}&=\begin{bmatrix}
     \Q_\sDesired       & \Zero_{2 \times 2}\\
     \Zero_{2 \times 2} & \text{Diag}\left[k_{\sI 1}, k_{\sI 2} \right]
     \end{bmatrix}
\end{align}
\end{subequations}
For the IA design, we then again neglect the interaction $( \vd=\vec{I}^\sDQ_\sDisturbance=\zero_2)$ in \eqref{eq:control:ia:closed_loop_phs_z_coord:dynamics}. 
Afterwards, we establish the assignment of the unactuated states (cf.\ \cite[(10)]{Donaire09})
\begin{equation} \label{eq:control:ia:trafo_unactuated}
   \Zstates_\sUnactuated \coloneqq \x_\sUnactuated 
\end{equation}
such that the equilibrium $\Zstates_\sUnactuated^*$ matches the desired one $\x_\sUnactuated^*$ implying $\VdqRef$ (see \eqref{eq:control:DGU_equilibrium_desired}, \eqref{eq:control:ia:equilibrium}). Subsequently, we find the transformation of the actuated states, which satisfies requirement \eqref{eq:control:ia:trafo_unactuated}, by solving (cf.\ \cite[Eq.~(12)]{Donaire09})
\begin{align} \label{eq:control:ia:trafo_actuated_equation}
\dot{\Zstates}_\sUnactuated&= 
\begin{bmatrix}
\frac{\Zstate_{\sActuated 1}- \bar{\Zstate}_{\sActuated 1}}{\Lt} + \freqRef \Ct \frac{\Zstate_{\sUnactuated 2}- \Zstate^*_{\sUnactuated 2}}{\Ct} - k_{\sI 1}\frac{\Zstate_{\sExtended 1}}{k_{\sI 1}}\\
\frac{\Zstate_{\sActuated 2}- \bar{\Zstate}_{\sActuated 2}}{\Lt} - \freqRef \Ct \frac{\Zstate_{\sUnactuated 2}- \Zstate^*_{\sUnactuated 2}}{\Ct} - k_{\sI 2}\frac{\Zstate_{\sExtended 2}}{k_{\sI 2}}
\end{bmatrix}
 = \dot{\x}_\sUnactuated \dots\nonumber\\
\dots&=
\begin{bmatrix}
\nu_{11}\frac{x_{\sActuated 1}- \bar{x}_{\sActuated 1}}{ \nu_{11}\Lt} + \freqRef \Ct \frac{x_{\sUnactuated 2}- x^*_{\sUnactuated 2}}{\Ct} \\
\nu_{22}\frac{x_{\sActuated 2}- \bar{x}_{\sActuated 2}}{\nu_{22}\Lt} - \freqRef \Ct \frac{x_{\sUnactuated 1}- x^*_{\sUnactuated 1}}{\Ct}
\end{bmatrix}
\end{align}
for $\Zstates_\sActuated \coloneqq \matrix{T}(\x,\Zstates_\sExtended)$, which yields
\begin{equation} \label{eq:control:ia:trafo_actuated}
\matrix{T}(\x,\Zstates_\sExtended)=\begin{bmatrix}
x_{\sActuated 1} - \bar{x}_{\sActuated 1} + \bar{\Zstate}_{\sActuated 1} + \Lt \Zstate_{\sExtended 1} \\
x_{\sActuated 2} - \bar{x}_{\sActuated 2} + \bar{\Zstate}_{\sActuated 2} + \Lt \Zstate_{\sExtended 2}
\end{bmatrix}
\end{equation}
Finally, we compute the integral control law \eqref{eq:control:ia:u_IA} from (cf.\ \cite[Eq.~(13)]{Donaire09}
\begin{align}
    \dot{\Zstates}_\sActuated &=\frac{\text{d} \matrix{T}(\x,\Zstates_\sExtended)}{\text{d} t}\\
    \begin{bmatrix}
    \alpha_{11} \frac{\Zstate_{\sActuated 1}- \bar{\Zstate}_{\sActuated 1}}{\Lt} - \nu_{11} \frac{\Zstate_{\sUnactuated 1}- \Zstate^*_{\sUnactuated 1}}{\Ct} \\
    \alpha_{22} \frac{\Zstate_{\sActuated 2}- \bar{\Zstate}_{\sActuated 2}}{\Lt} - \nu_{22} \frac{\Zstate_{\sUnactuated 2}- \Zstate^*_{\sUnactuated 2}}{\Ct}
\end{bmatrix} &=
\begin{bmatrix}
\dot{x}_{\sActuated 1} + \Lt \dot{\Zstate}_{\sExtended 1}\\
\dot{x}_{\sActuated 2} + \Lt \dot{\Zstate}_{\sExtended 2}\\
\end{bmatrix}
\end{align}
by inserting $\dot{\x}_{\sActuated}$ from \eqref{eq:control:ia:closed_loop_phs}, as well as \eqref{eq:control:ia:integrator_states}, \eqref{eq:control:ia:trafo_unactuated}, \eqref{eq:control:ia:trafo_actuated}, and solving for $\vu_\sIA$.
In order to investigate the equilibrium established with \eqref{eq:control:ia:u_IA} under piecewise constant interactions $\bar{\vd}=-\bar{\vec{I}}^\sDQ_\sDisturbance$, we rewrite \eqref{eq:control:ia:closed_loop_phs_z_coord} by including $\vd=\bar{\vd}$ in \eqref{eq:control:ia:closed_loop_phs_z_coord:dynamics} into the Hamiltonian as  
 %
%
\begin{align} \label{eq:ia:rewrite_z_Hamil_stability}
    \Hamil_{\sDesired \sZ}(\widetilde{\Zstates}) &= \frac{1}{2} \underbrace{\begin{bmatrix}
     \Zstates_\sActuated - \bar{\Zstates}_\sActuated\\
     \Zstates_\sUnactuated - \Zstates_\sUnactuated^*\\
     \Zstates_\sExtended -\bar{\vd}
     \end{bmatrix}^\Transpose}_{\widetilde{\Zstates}^\Transpose}
     \Q_{\sDesired \sZ}
     \underbrace{\begin{bmatrix}
     \Zstates_\sActuated - \bar{\Zstates}_\sActuated\\
     \Zstates_\sUnactuated - \Zstates_\sUnactuated^*\\
     \Zstates_\sExtended - \bar{\vd}
     \end{bmatrix}}_{\widetilde{\Zstates}}
\end{align}
For the equilibrium in $\Zstate$-coordinates $\bar{\Zstates}^*=\left[\bar{\Zstates}_\sActuated, \Zstates_\sUnactuated^*, \bar{\Zstates}_\sExtended \right]^\Transpose$ follows from \eqref{eq:control:ia:trafo_unactuated} and \eqref{eq:ia:rewrite_z_Hamil_stability} that 
\begin{align} 
\Zstates^*_\sUnactuated&=\x^*_\sUnactuated=\Ct \VdqRef \label{eq:control:ia:x_nu_equilibrium}\\ \bar{\Zstates}_\sExtended&=\bar{\vd}=-\bar{\vec{I}}^\sDQ_\sDisturbance. \label{eq:control:ia:z_e_equilibrium}
\end{align}
For $\lim_{t \to \infty} \Zstates_\sActuated = \bar{\Zstates}_\sActuated$ to hold, it follows from \eqref{eq:control:ia:trafo_actuated} with \eqref{eq:control:design:current_equilibrium} and \eqref{eq:control:ia:z_e_equilibrium} that
\begin{equation} \label{eq:control:ia:x_alpha_equilibrium}
    \lim_{t \to \infty} \! \x_\sActuated= \xRef_\sActuated - \Lt \bar{\Zstates}_\sExtended=\!\!
    \begin{bmatrix}
     \Lt \left(-\freqRef \Ct \VqRef + \bar{I}^\sD_\sDisturbance \right)\\
    \Lt \left(\freqRef \Ct \VdRef + \bar{I}^\sQ_\sDisturbance\right)
    \end{bmatrix} \!\eqqcolon \bar{\x}_\sActuated^*,
\end{equation}
With \eqref{eq:control:ia:x_nu_equilibrium} and \eqref{eq:control:ia:x_alpha_equilibrium} we obtain the equilibrium \eqref{eq:control:ia:equilibrium} in $x$-coordinates.
\IEEEQEDclosed

The overall control input \eqref{eq:control:ia:general} is now calculated with \eqref{eq:control:result:u_simple} and \eqref{eq:control:ia:u_IA} to
\begin{equation} \label{eq:control:overall_control_law}
\hspace{-0.8em}	\underset{(2\times 1)}{\vu(\x)} \, {=} \left[ \begin{array}{@{\,}c@{\,}}
		\frac{\alpha_{11}}{\nu_{11}} (\Id+\freqRef \Ct \VqRef)\\ 
		- \nu_{11} (\Vd - \VdRef) + \Rt \Id - \freqRef \Lt \Iq + \Vd \\
		-\alpha_{11} k_{\sI 1} \int (\VdRef- \Vd) \text{d} t + k_{\sI 1} \Lt (\VdRef - \Vd)	\Big.	\\
		\hdashline
		\Big. \frac{\alpha_{22}}{\nu_{22}} (\Iq-\freqRef \Ct \VdRef)  \\ 
		- \nu_{22} (\Vq - \VqRef) + \Rt \Iq + \freqRef \Lt \Id + \Vq\\
		- \alpha_{22} k_{\sI 2} \int (\VqRef- \Vq) \text{d} t + k_{\sI 2} \Lt (\VqRef - \Vq)
	\end{array} \right]
\end{equation}
with control parameters $ \alpha_{11}, \alpha_{22} <0$, $\nu_{11}, \nu_{22} > 0 $, and $k_{\sI1}, k_{\sI2} >0$.
\begin{remark}
Note that the IA \eqref{eq:control:ia:u_IA} does not affect the passivity-related voltage stability property of the closed-loop DGU model \eqref{eq:control:ia:closed_loop_phs}. It only establishes a new current equilibrium  $\bar{\x}_\sActuated^*=\bar{\vec{I}}^{\sDQ *}_\sInput= \left[  -\freqRef \Ct \VqRef + \bar{I}^\sD_\sDisturbance, \,  \freqRef \Ct \VdRef + \bar{I}^\sQ_\sDisturbance \right]^\Transpose$ depending on the steady-state interaction current $\bar{\vec{I}}^\sDQ_\sDisturbance$, while preserving the desired voltage equilibirum $\x_\sUnactuated^*=\Ct \VdqRef$ and the assigned port-Hamiltonian structure, i.e.\ $\Jd, \Rd, \Hd(\x)$, albeit in new $\Zstate$-coordinates.
\end{remark}

\subsection{Passivity-Based PI Control Design} \label{control:pulkit}
To guarantee voltage and frequency in ImGs, one can also equip DGUs \eqref{eq:modelling:dgu_phs} with multi-variable integral controllers, as those proposed in \cite{Nahata19ecc}.
To be in line with \cite{Nahata19ecc}, we first permutate the state vector such that $\x_i = \left[ \Cti \Vd_i, \Cti \Vq_i, \Lti \Itd[i], \Lti \Itq[i] \right]^\Transpose, i\in\DD_\sPI,$ and introduce the integrator dynamics
\begin{equation}
\dot{\vec{v}}_{i} = \VdqRef_i-\z_{i},\quad i\in\DD_\sPI
\label{eq:intdynamics}
\end{equation}
Subsequently, we equip DGU \eqref{eq:modelling:dgu_phs} with a state-feedback controller
\begin{equation} \label{eq:control:state_feedback}
{\vu}_{i}=\matrix{\KK}_{i}{\hat{\x}}_{i}, \quad i\in\DD_\sPI
\end{equation}
where $\hat{\x}_{i}=[{\x}^\Transpose_{i},\vec{v}_{i}^\Transpose]^\Transpose$ is the augmented, permutated state vector and $\matrix{\KK}_{i}=\left[\nicefrac{\KK_{11,i}}{\Cti}, \nicefrac{\KK_{12,i}}{\Lti} , \KK_{13,i}\right]=\Reals^{2\times 6}$ is the matrix collecting the control gains. Let $\hat{\bar{\x}}^*_{i}=[{\xRef}_{i}^{*\Transpose},{\vec{v}^*_{i}}^\Transpose]^\Transpose$ denote the equilibrium of DGU $i\in\DD_\sPI$ controlled with \eqref{eq:control:state_feedback}. Using the change of variables $\tilde{\x}_i={\hat{\x}}_{i}-\hat{\xRef}^*_{i}$, we write the shifted, augmented model of the PI-controlled DGU as (cf.\ \cite[Eq.~(17)-(20)]{Nahata19ecc})
\begin{equation}
\label{eq:control:dgu_augemented_shifted}
\begin{aligned}
\dot{\xtilde}_i &= \matrix{F}\xtilde_i+\tilde{\K}_{i}\left(\vd_i - \bar{\vd}_i \right)\\
\z_i-\z^*_i     &= \Vdq_i -\VdqRef_i
\end{aligned},~i\in\DD_\sPI
\end{equation}
where
\begin{align}
\label{eq:control:dgu_pulkit_submatrices}
	\matrix{F}_{i}&\!=\!\!\!
    \left[\!\!\! \begin{array}{ccc}
	\mathcal{\tilde{\matrix{A}}}_{11,i} &  \frac{1}{\Lti}\One_{2\times2} & \Zero_{2\times 2} \\
	-\frac{1}{\Cti}\left(\One_{2\times 2}- \matrix{\KK}_{11,i}\right) &\mathcal{\tilde{\matrix{A}}}_{22,i}+\frac{\matrix{\KK}_{12,i}}{\Lti} &\matrix{\KK}_{13,i}\\
	-\frac{1}{\Cti}\One_{2\times 2} &\Zero_{2\times 2} & \Zero_{2\times 2} \\
	\end{array}\!\!\!
	\right]\\
	\mathcal{\tilde{\matrix{A}}}_{11,i}&=\!\!\begin{bmatrix}
	0 & \omega_0\\
	-\omega_0 & 0
    \end{bmatrix},
    \mathcal{\tilde{\matrix{A}}}_{22,i}=\!\!\begin{bmatrix}
    -\frac{\Rti}{\Lti} & \omega_0\\
    -\omega_0 & -\frac{\Rti}{\Lti}
    \end{bmatrix},
    \tilde{\K}_i=\!\!\begin{bmatrix}
    \One_{2\times 2}\\
    \Zero_{2\times 2}\\
    \Zero_{2\times 2}
    \end{bmatrix} \nonumber
\end{align}
\section{Modular Stability Analysis} \label{sec:stability}

With the control design completed, we are now able to address Problem~\ref{problem} by a modular, bottom-up stability analysis of the ImG equilibrium $\xRef_\sMG^*$. Following the lines of Lemma~\ref{lemma:modular_stability}, \cite[Theorem 1]{Nahata20}, and \cite[Proposition 1]{Strehle20ifac},  $\xRef_\sMG$ is \emph{stable} if two conditions are satisfied:
\begin{enumerate}
\item[C.1] the ImG subsystems---closed-loop DGU \eqref{eq:control:ia:closed_loop_phs_z_coord} and \eqref{eq:control:dgu_augemented_shifted}, respectively, load nodes \eqref{eq:modelling:loadphs}, and power lines \eqref{eq:modelling:line_phs}---are connected in a skew-symmetric, i.e.~ power-preserving, fashion defined by \eqref{eq:prems:skew_symm_inter};
\item[C.2] each subsystem is EIP with positive definite storage function $S_i(\x_i,\xRef_i) \succ 0$.
\end{enumerate}
In Section~\ref{sec:stability:intercon}, we thus analyze the interconnection structure of the ImG and show that C.1 holds by construction. In Sections~\ref{sec:stability:eip_dgu}--\ref{sec:stability:eip_line}, we then analyze \eqref{eq:control:ia:closed_loop_phs_z_coord} respectively \eqref{eq:control:dgu_augemented_shifted}, \eqref{eq:modelling:loadphs}, and \eqref{eq:modelling:line_phs} regarding EIP. 
We show that if a controller results in a closed-loop DGU that satisfies both C.1 and C.2, it can seamlessly connect to the ImG without spoiling voltage and frequency stability.
\emph{Asymptotic stability} of $\xRef_\sMG^*$ and thus $\bar{\vec{V}}^{\sDQ*}_k, k\in\MM$ follows from LaSalle's Invariance principle as illustrated in Section~\ref{sec:stability:img}, Theorem~\ref{theorem:ImG_stability}. We conclude our analysis by investigating the impact of actuator saturation on the stability of the ImG equilibrium $\xRef_\sMG^*$ in Section~\ref{sec:stability:saturation}.

%
%
\subsection{Interconnection Structure} \label{sec:stability:intercon}
 \begin{lemma}
\label{lem:ImG_skew_sym_inter}
The electrical interconnections in ImGs between controlled DGUs and load nodes, respectively, and power lines given by \eqref{eq:inputDGU}, \eqref{eq:inputLoad}, and \eqref{eq:inputLines} are skew-symmetric. 
\end{lemma}
\IEEEproof
Consider the inputs \eqref{eq:inputDGU} and \eqref{eq:inputLoad} to the closed-loop DGU $i\in\DD$ (e.g.\ in \eqref{eq:control:ia:closed_loop_phs_z_coord:dynamics})\footnote{Note that the interaction (coupling) port $\vd_i, \z_i$ is not affected by control and thus remains the same for DGUs controlled with different controllers.}, and load node $j$ in \eqref{eq:modelling:loadphs}, respectively. With
\eqref{eq:modelling:line_phs_z}, \eqref{eq:inputDGU} and \eqref{eq:inputLoad} can equivalently be written as
\begin{equation} \label{eq:stability:inputDGULoad}
\vd_k=\sum_{l \in \mathcal{N}^+_k}\underbrace{\left[ -\One_{2 \times 2} \; \Zero_{2 \times 2} \right]}_{\matrix{\phi}_{kl}}\z_l+\sum_{l \in \mathcal{N}^-_k}\underbrace{\left[\Zero_{2 \times 2} \; -\One_{2 \times 2} \right]}_{\matrix{\phi}_{kl}=-\matrix{\phi}_{lk}^\Transpose}\z_l
\end{equation}
%
%
%
for $k\in\MM$. With \eqref{eq:modelling:dgu_phs_z} and \eqref{eq:modelling:loadphs_z}, input \eqref{eq:inputLines} of line $l \in \PP$ becomes
\begin{equation} \label{eq:stability:inputLine}
    \vd_{\sLine} = 
    \sum_{k \in \mathcal{N}_l^+}
    \underbrace{\begin{bmatrix}
    \Zero_{2 \times 2}\\
    \One_{2 \times 2}
    \end{bmatrix}}_{\matrix{\phi}_{lk}} \z_k +
    \sum_{k \in \mathcal{N}_l^-}
    \underbrace{\begin{bmatrix}
    \One_{2 \times 2}\\
    \Zero_{2 \times 2}
    \end{bmatrix}}_{\matrix{\phi}_{lk}= -\matrix{\phi}_{kl}^\Transpose} \z_k, \; l \in \PP
\end{equation}
%
Note that if $k \in \NN^+_l$ in \eqref{eq:stability:inputLine}, then $l \in \NN^-_k$ in \eqref{eq:stability:inputDGULoad} and $\matrix{\phi}_{kl}=\left[\Zero_{2 \times 2} \; -\One_{2 \times 2} \right]=-\matrix{\phi}_{lk}^\Transpose$. Conversely,
if $l \in \NN^+_k$ in \eqref{eq:stability:inputDGULoad}, then $k \in \NN^-_l$ in \eqref{eq:stability:inputLine} and $\matrix{\phi}_{lk}=\left[\One_{2 \times 2} \; \Zero_{2 \times 2} \right]^\Transpose=-\matrix{\phi}_{kl}^\Transpose$.
Combining \eqref{eq:stability:inputDGULoad} and \eqref{eq:stability:inputLine} yields the overall interconnection structure in the desired skew-symmetric form \eqref{eq:prems:skew_symm_inter} with $i=1,\dots,|\VV|=|\MM|\cup|\PP|$. \IEEEQEDclosed



%
\subsection{Equilibrium-Independent Passivity of DGUs controlled with Port-Hamiltonian-based Controllers} \label{sec:stability:eip_dgu}
The overall closed-loop DGU ISO-PHS model, resulting from equipping \eqref{eq:modelling:dgu_phs} with the IDA-PBC + IA control law in \eqref{eq:control:overall_control_law}, is given by \eqref{eq:control:ia:closed_loop_phs_z_coord} with $i\in\DD_\sPHS$. 
In order to investigate its EIP, we rewrite \eqref{eq:control:ia:closed_loop_phs_z_coord} in new $\xtilde$-coordinates as
\begin{subequations} \label{eq:stability:dgu:closed_loop_phs_xtilde}
    \begin{align} \label{eq:stability:dgu:closed_loop_phs_xtilde:dynamics}
    \xtildedot_i &=
    \left[ \J_{\sDesired\sZ,i}-\R_{\sDesired\sZ,i} \right] \frac{\partial \Hamil_{i}(\xtilde_i)}{\partial \xtilde_i}+\begin{bmatrix}
    \K_i\\
    \Zero_{2\times2}
    \end{bmatrix} ({\vd}_i-\bar{\vd}_i)\\
	{\z_{i}-\z_{i}^*} &= \begin{bmatrix}
    \K_i\\
    \Zero_{2\times2}
    \end{bmatrix}^\Transpose \frac{\partial \Hamil_{i}(\xtilde)}{\partial \xtilde}, \quad i\in\DD_\sPHS\\
    \label{eq:stability:dgu:closed_loop_phs_xtilde:hamil}
     \Hamil_{i}(\xtilde_i) &= \frac{1}{2} \xtilde_{i}^\Transpose \Q_{\sDesired \sZ,i} \xtilde_{i}
\end{align}
\end{subequations}
with $\xtilde_i= \Zstates_i-\bar{\Zstates}_i^*$, $\bar{\Zstates}_i^*=\left[\bar{\Zstates}_{\sActuated,i}, \Zstates_{\sUnactuated,i}^*, \bar{\Zstates}_{\sExtended,i} \right]^\Transpose$, $\frac{\partial \Hamil_{i}(\xtilde_i)}{\partial \xtilde_i}=\left[ \frac{\Zstate_{\sActuated 1} - \bar{{\Zstate}}_{\sActuated 1}}{\Lti \nu_{11,i}}, \frac{\Zstate_{\sActuated 2} - \bar{{\Zstate}}_{\sActuated 2}}{\Lti \nu_{22,i}}, \frac{\Zstate_{\sUnactuated 1} - \Zstate^*_{\sUnactuated 1}}{\Cti}, \frac{\Zstate_{\sUnactuated 2} - \Zstate^*_{\sUnactuated 2}}{\Cti}, \frac{\Zstate_{\sExtended1,i}-\bar{\Zstate}_{\sExtended1,i}}{k_{\sI 1,i}}, \frac{x_{\sExtended 2,i}-\bar{\Zstate}_{\sExtended2,i}}{k_{\sI 2,i}}  \right]^\Transpose$.
The time-derivative of the Hamiltonian \eqref{eq:stability:dgu:closed_loop_phs_xtilde:hamil} is
\begin{align} \label{eq:stability:dgu:hamil_dt}
\dot{\Hamil}_{i}(\xtilde_i)=& -\underbrace{\frac{\partial^\Transpose \Hamil_{i}(\xtilde_i)}{\partial \xtilde_i} \R_{\sDesired\sZ,i} \frac{\partial \Hamil_{i}(\xtilde_i)}{\partial \xtilde_i}}_{\eqqcolon \psi_{i}(\xtilde_i)}\\
&+ ({\z_{i}-\z_{i}^*})^\Transpose
	({\vd}_i-\bar{\vd}_i) \nonumber \\
&\leq ({\z_{i}-\z_{i}^*})^\Transpose
	({\vd}_i-\bar{\vd}_i) , \quad \forall \xtilde_i \neq \zero_2, \nonumber
\end{align}
with $\psi_{i}(\xtilde_i)\geq 0$ as $\R_{\sDesired\sZ,i} \succeq 0$ (cf.\ \eqref{eq:control:ia:Rcz}). Thus,  \eqref{eq:stability:dgu:closed_loop_phs_xtilde} is passive, but not strictly passive, with respect to the shifted supply rate $({\vd}_i-\bar{\vd}_i)^\Transpose ({\z_{i}-\z_{i}^*})$ and the positive definite storage function \eqref{eq:stability:dgu:closed_loop_phs_xtilde:hamil} (cf.\ \eqref{eq:control:ia:Qcz}). According to Defintion~\ref{def:eip}, \eqref{eq:control:ia:closed_loop_phs_z_coord} is thus merely EIP. The same result can be obtained from Lemma~\ref{lemma:eip_systems} by noting that \eqref{eq:control:ia:closed_loop_phs_z_coord} is a linear ISO-PHS with quadratic, positive definite Hamiltonian. 

\subsection{Equilibrium-Independent Passivity of PI-controlled DGUs}\label{sec:stability:eip_dgu_pulkit}
In addition to \cite{Nahata19ecc}, where only passivity is considered, we prove that equipping DGUs \eqref{eq:modelling:dgu_phs} with the multi-variable PI controllers from \cite{Nahata19ecc} also renders the closed-loop DGU EIP. For this we consider \eqref{eq:control:dgu_augemented_shifted} and propose the candidate storage function
\begin{equation}
\label{eq:stability:pulkit_Lyapunovfunction}
\Hamil_{i}(\tilde{\x}_{i})=\dfrac{1}{2}{\tilde{\x}_{i}}^T\vec{P}_i{\tilde{\x}}_{i}, \quad i\in \DD_\sPI
\end{equation}
where $\matrix{P}_{i}=\matrix{P}^T_{i} \in \Reals^{6 \times 6}$, a positive definite matrix, has the following structure
\begin{equation}
	\matrix{P}_{i}=\left[ \begin{array}{c|cc}
 	\One_{2\times 2} & \Zero_{4\times 4} \\
 	\hline
 	\Zero_{4\times 4} & \matrix{P}_{22,i}
 	\end{array}\right], \quad i\in \DD_\sPI
\end{equation}
with $\matrix{P}_{22,i} \succ 0$. The time-derivative of \eqref{eq:stability:pulkit_Lyapunovfunction} is
\begin{equation} \label{eq:stability:pulkit:hamil_dt}
   \dot{\Hamil}_{i}(\xtilde_i)=\underbrace{\tilde{\x}^\Transpose \matrix{\QQ}_i \tilde{\x}}_{_{\eqqcolon \psi_i(\xtilde_i)}}+ ({\z_{i}-\z_{i}^*})^\Transpose
	({\vd}_i-\bar{\vd}_i), \quad i\in \DD_\sPI
\end{equation}
where $\matrix{\QQ}_i=\matrix{F}_i^\Transpose \matrix{P}_i+ \matrix{P}_i \matrix{F}_i$. Following \cite{Nahata19ecc}, the control gains $\matrix{\KK}_i$ in $\matrix{F}_i$ (see \eqref{eq:control:dgu_pulkit_submatrices} can be chosen such that $\psi_i(\xtilde_i)\geq0$, making the DGU EIP. For further details regarding the calculation of the control gains, the reader is deferred to \cite{Nahata19ecc}.

\subsection{Equilibrium-Independent Passivity of Load Nodes} \label{sec:stability:eip_load}
Consider lone-standing load nodes $j \in \LL$ described by the ISO-PHS load model \eqref{eq:modelling:loadphs}. By rewriting \eqref{eq:modelling:loadphs} with respect to any feasible, unknown equilibrium $\xLRef=C_\sLoad \bar{\vec{V}}^\sDQ_\sLoad, j \in \LL$, we obtain
\begin{subequations} \label{eq:load_passivity:load_errorsystem}
	\begin{align}
		\xtilde_\sLoad&=
		\JL
		\frac{\partial \Hamil_\sLoad(\xtilde_\sLoad)}{\partial \xtilde_\sLoad}
		- \left( \RL(\xtilde_\sLoad+\xRef_\sLoad)-\RL(\bar{\x}_\sLoad)\right) \nonumber\\
		&\dots +\KL \left( \vd_\sLoad-\bar{\vd}_\sLoad\right),\\
		\z_\sLoad-\zRef_\sLoad&=\KL^\Transpose \frac{\partial \Hamil_\sLoad(\xtilde_\sLoad)}{\partial \xtilde_\sLoad}, \label{eq:load_passivity:load_errorsystem:shifted_output}\\
		\label{eq:load_passivity:load_errorsystem:Hamiltonian}
		\Hamil_\sLoad(\xtilde_\sLoad) &= \frac{1}{2} \, \xtilde_\sLoad^\Transpose \; \text{Diag}\left[\frac{1}{C_\sLoad}, \frac{1}{C_\sLoad} \right] \xtilde_\sLoad
	\end{align}
\end{subequations}
with $\xtilde_\sLoad=\xL-\xRef_\sLoad, \frac{\partial \Hamil_\sLoad(\xtilde_\sLoad)}{\partial \xtilde_\sLoad}=\VLdq - \bar{\vec{V}}^\sDQ_\sLoad, \vd_\sLoad-\bar{\vd}_\sLoad=\Izdq[\sLoad]-\bar{\vec{I}}^\sDQ_{\sDisturbance \sLoad}, \z_\sLoad-\zRef_\sLoad= \VLdq - \bar{\vec{V}}^\sDQ_\sLoad$.
The original load model \eqref{eq:modelling:loadphs} is strictly EIP, iff the load model error system \eqref{eq:load_passivity:load_errorsystem} is strictly passive with respect to the shifted supply rate $(\z_\sLoad-\zRef_\sLoad)^\Transpose (\vd_\sLoad-\bar{\vd}_\sLoad)$ and the positive definite storage function \eqref{eq:load_passivity:load_errorsystem:Hamiltonian}, i.e. $\Hamil_\sLoad(\xtilde_\sLoad): \Reals^2 \to \Reals_{\geq 0}, \Hamil_\sLoad(\zero_2)=0, \Hamil_\sLoad(\xtilde_\sLoad)>0  \, \forall  \xtilde_\sLoad\neq \zero_2$. This is the case, if
%
\begin{align}\label{eq:load_passivity:passivity_ineq}
	\dot{\Hamil}_\sLoad(\xtilde_\sLoad)=&-\underbrace{\frac{\partial^\Transpose  \Hamil_\sLoad(\xtilde_\sLoad)}{\partial \xtilde_\sLoad} \left( \RL(\xtilde_\sLoad+\xRef_\sLoad)-\RL(\bar{\x}_\sLoad)\right)}_{\eqqcolon \psi_\sLoad(\xtilde_\sLoad)} \nonumber\\
	&+ (\z_\sLoad-\zRef_\sLoad)^\Transpose (\vd_\sLoad-\bar{\vd}_\sLoad)\\
	&< (\z_\sLoad-\zRef_\sLoad)^\Transpose (\vd_\sLoad-\bar{\vd}_\sLoad) \quad \forall \xtilde_\sLoad \neq \zero_2 \nonumber
\end{align}
holds. By crossing out the supply rates and substituting
$\frac{\partial^\Transpose  \Hamil_\sLoad(\xtilde_\sLoad)}{\partial \xtilde_\sLoad}=\left(\frac{\xtilde_\sLoad}{C_\sLoad}\right)^\Transpose =\left( \VLdq-\bar{\vec{V}}^\sDQ_\sLoad \right)^\Transpose$, \eqref{eq:load_passivity:passivity_ineq} results in
\begin{align}\label{eq:load_passivity:monotonicity_ineq_states}
	  \!\! \left(\frac{\xtilde_\sLoad}{C_\sLoad}\right)^\Transpose \!\! \left(\RL(\xtilde_\sLoad+\xRef_\sLoad)-\RL(\bar{\x}_\sLoad)\right)&>0, \, \forall \xtilde_\sLoad \neq \zero_2, \\
	  \label{eq:load_passivity:monotonicity_ineq}
	 \!\! \left( \VLdq-\bar{\vec{V}}^\sDQ_\sLoad \right)^\Transpose \!\!  \left( \ILdq(\VLdq)-\ILdq(\bar{\vec{V}}^\sDQ_\sLoad)\right)&>0, \, \forall \xtilde_\sLoad \neq \zero_2.
\end{align}
In the case of lone-standing load nodes $j \in \LL$ comprising ZIP and EXP loads \eqref{eq:modelling:nonlinear_R_zip_exp}, we conducted a comprehensive study in \cite{Strehle20ifac} yielding sufficient conditions for \eqref{eq:load_passivity:monotonicity_ineq} and thus \emph{strict EIP}. 
In order to present a consistent framework, we shortly recall the main result of \cite{Strehle20ifac}.
\begin{proposition} \label{prop:passivity_conditions:load}
The ISO-PHS load model \eqref{eq:modelling:loadphs} of any lone-standing load node $j \in \LL$ is \emph{strictly EIP}, i.e. $\psi_\sLoad(\xtilde_\sLoad) \succ0 $,
\begin{itemize}
	\item  if for ZIP loads \eqref{eq:modelling:nonlinear_R_zip_exp} it holds that
\end{itemize}
	\begin{subequations} \label{eq:load_passivity:zip_restrictions}
			\begin{align} \label{eq:load_passivity:zip_restrictions:first}
			\lYp[\sLoad] + \dfrac{\lIp[\sLoad]}{2 \Vamp[\sLoad]} &> 0,\\
			\lYp[\sLoad]^2  \Vamp[\sLoad]^4 + \lYp[\sLoad] \lIp[\sLoad] \Vamp[\sLoad]^3 &> \frac{1}{4} \lIq[\sLoad] \Vamp[\sLoad]^2 + \left(\lIp[\sLoad] \lPp[\sLoad] + \lIq[\sLoad] \lPq[\sLoad]\right) \Vamp[\sLoad] \dots \nonumber \\
			& \; \dots + \left(\lPp[\sLoad]^2 +\lPq[\sLoad]^2\right). \label{eq:load_passivity:zip_restrictions:second}
		\end{align}
	\end{subequations}
\begin{itemize}
	\item if for EXP loads \eqref{eq:modelling:nonlinear_R_zip_exp}  it holds that 
	\end{itemize}
	\begin{subequations} \label{eq:load_passivity:exp_restrictions}
		\begin{align} \label{eq:load_passivity:exp_restrictions:first}
			\!\!\lNp[\sLoad] \lActNom[\sLoad] &> 0 , \\ 
			\label{eq:load_passivity:exp_restrictions:second}
			\!\!4\left(\lNp[\sLoad] - 1\right) \lActNom[\sLoad]^2 \left(\frac{\Vamp[\sLoad]}{\VampNominal}\right)^{2 \lNp[\sLoad]}\!\!\!\! &> \! \left(\lNq[\sLoad] - 2\right)^2 \! \lReacNom[\sLoad]^2 \! \left(\frac{\Vamp[\sLoad]}{\VampNominal}\right)^{2 \lNq[\sLoad]} 
		\end{align}
	\end{subequations}
\end{proposition}
%
%
\begin{remark} \label{remark:dH_dt_load_at_dgu}
Following Remark~\ref{remark:load_at_DGU}, we may include a local ZIP or EXP load \eqref{eq:modelling:nonlinear_R_zip_exp} at any closed-loop DGU (e.g.\ in \eqref{eq:stability:dgu:closed_loop_phs_xtilde:dynamics} (resp.\ \eqref{eq:control:dgu_augemented_shifted})) by subtracting
    $\begin{bmatrix}
    \K_i\\
    \Zero_{2\times2}
    \end{bmatrix} \left(\RR_i(\xtilde_i +\bar{\Zstates}^*_i)- \RR_i(\bar{\Zstates}^*_i)\right)$.
Subsequently, \eqref{eq:stability:dgu:hamil_dt} (resp.\ \eqref{eq:stability:pulkit:hamil_dt}) becomes
\begin{align} \label{eq:stability:dgu_load:hamil_dt}
\dot{\Hamil}_{i}(\xtilde_i)=& -\psi_{i}(\xtilde_i) -
\psi_{\RR_i}(\xtilde_i) \\
&+ ({\z_{i}-\z_{i}^*})^\Transpose
	({\vd}_i-\bar{\vd}_i) \nonumber \\
&\leq ({\z_{i}-\z_{i}^*})^\Transpose
	({\vd}_i-\bar{\vd}_i) , \quad \forall \xtilde_i \neq \zero_2, \nonumber
\end{align}
with
\begin{equation}
    \psi_{\RR_i}(\xtilde_i) \coloneqq \frac{\partial^\Transpose \Hamil_{i}(\xtilde_i)}{\partial \xtilde_i} \begin{bmatrix}
    \K_i\\
    \Zero_{2\times2}
    \end{bmatrix} \left(\RR_i(\xtilde_i +\bar{\Zstates}^*_i)- \RR_i(\bar{\Zstates}^*_i)\right).
\end{equation}
Note that unlike in \eqref{eq:load_passivity:passivity_ineq}--\eqref{eq:load_passivity:monotonicity_ineq}, not all states $\xtilde_i$ are affected by  $\psi_{\RR_i}(\xtilde_i)$ and thus only $\psi_{\RR_i}(\xtilde_i) \geq 0$ under the load EIP conditions of Proposition~\ref{prop:passivity_conditions:load}. Furthermore, $\psi_{i}(\xtilde_i) + \psi_{\RR_i}(\xtilde_i) \geq 0$ and the overall closed-loop DGU \eqref{eq:stability:dgu:closed_loop_phs_xtilde} (resp.\ \eqref{eq:control:dgu_augemented_shifted}) supplying a local ZIP or EXP load \eqref{eq:modelling:nonlinear_R_zip_exp} remains merely EIP.
\end{remark}

\subsection{Equilibrium-Independent Passivity of Lines} \label{sec:stability:eip_line}
As the ISO-PHS \eqref{eq:modelling:line_phs} of the $\Pi$-model power line is linear with positive definite dissipation matrix \eqref{eq:modelling:line_JR}, it is \emph{strictly EIP} according to Lemma~\ref{lemma:eip_systems} (see also \cite[p.~136]{vdS17}). For the subsequent LaSalle analysis, we introduce the shifted Hamiltonian
\begin{equation} \label{eq:stability:line:hamil}
    	\Hamil_\sLine(\xtilde_\sLine) = \frac{1}{2} \, \xtilde_\sLine^\Transpose \; \text{Diag}\left[\frac{1}{L_\sLine}, \frac{1}{L_\sLine} \right] \xtilde_\sLine
\end{equation}
with $\xtilde_\sLine=\x_\sLine-\bar{\x}_\sLine$, $\bar{\x}_\sLine=L_\sLine \bar{\vec{I}}^\sDQ_\sLine$. The time-derivative of \eqref{eq:stability:line:hamil} is
\begin{align} \label{eq:stability:line:hamil_dt}
    \dot{\Hamil}_\sLine(\xtilde_\sLine)=&-\underbrace{\frac{\partial^\Transpose \Hamil_\sLine(\xtilde_\sLine)}{\partial \xtilde_\sLine} \R_\sLine \frac{\partial \Hamil_\sLine(\xtilde_\sLine)}{\partial \xtilde_\sLine}}_{\eqqcolon \psi_\sLine(\xtilde_\sLine)}+ \left(\z_\sLine - \bar{\z}_\sLine \right)^\Transpose \left(\vd_\sLine - \bar{\vd}^*_\sLine \right)\\
    &< \left(\z_\sLine - \bar{\z}_\sLine \right)^\Transpose \left(\vd_\sLine - \bar{\vd}^*_\sLine \right), \, \forall \xtilde_\sLine \neq \zero_2, \nonumber
\end{align}
with $\psi_\sLine(\xtilde_\sLine)>0$ for all $\xtilde_\sLine \neq \zero_2$ as $\R_\sLine\succ0$ (cf.\ \eqref{eq:modelling:line_JR}).

\subsection{Stability of a Modular Islanded AC Microgrid} \label{sec:stability:img}
With the EIP properties of closed-loop DGU, load node, and line models analyzed, we are now able to formulate our main results regarding stability of the overall ImG equilibrium $\xRef_\sMG$.
\begin{theorem} \label{theorem:ImG_stability}
Consider an ImG with arbitrary topology as described in Section~\ref{sec:modelling:img} consisting of $|\DD|=D$ closed-loop DGUs with optional local ZIP/EXP loads, $|\LL|=L$ nonlinear, lone-standing ZIP/EXP loads, and $|\PP|=P$ $\Pi$-model power lines described by
\eqref{eq:control:dgu_augemented_shifted} (resp.\ \eqref{eq:stability:dgu:closed_loop_phs_xtilde}), \eqref{eq:load_passivity:load_errorsystem}, and \eqref{eq:modelling:line_phs}, respectively. Under Proposition~\ref{prop:passivity_conditions:load}, the composite ImG is EIP and exhibits an asymptotically stable equilibrium
\begin{equation} \label{eq:stability:MG_equilibrium}
\bar{\x}^*_\sMG = \left[\xRef^*_1, \dots, \xRef^*_{L+D+P} \right]^\Transpose \in \XX_\sMG \subset \Reals^{2(3D+L+P)}
\end{equation}
\end{theorem}
\IEEEproof
%
%
As all subsystems are at least EIP with $\psi_i(\xtilde_i)\geq0$, $\psi_j(\xtilde_j)\succ0$ under Proposition~\ref{prop:passivity_conditions:load}, and $\psi_l(\xtilde_l)\succ 0$ for all $\x_\sMG \neq \xRef_\sMG$, and their interconnection is skew-symmetric, i.e.\ power-preserving, (see Lemma~\ref{lem:ImG_skew_sym_inter}), the composite ImG model is EIP.
Its composite storage function is given by the sum of the closed-loop DGU, load, and line Hamiltonians, respectively storage functions, \eqref{eq:stability:dgu:closed_loop_phs_xtilde:hamil} (resp.\ \eqref{eq:stability:pulkit_Lyapunovfunction}), \eqref{eq:load_passivity:load_errorsystem:Hamiltonian}, \eqref{eq:stability:line:hamil}
\begin{equation} \label{eq:stability:MG_hamil}
    \Hamil_\sMG(\xtilde_\sMG)=\sum_{i \in \DD} \Hamil_i(\xtilde_i) +\sum_{j \in \LL} \Hamil_\sLoad(\xtilde_\sLoad) +  \sum_{l \in \PP} \Hamil_\sLine(\xtilde_\sLine)
\end{equation}
with $\xtilde_\sMG=\left[\xtilde_1, \dots, \xtilde_{L+D+P} \right]^\Transpose$. As
\eqref{eq:stability:dgu:closed_loop_phs_xtilde:hamil} (resp.\ \eqref{eq:stability:pulkit_Lyapunovfunction}), \eqref{eq:load_passivity:load_errorsystem:Hamiltonian}, and \eqref{eq:stability:line:hamil} are positive definite, \eqref{eq:stability:MG_hamil} is positive definite.
%
%
With \eqref{eq:stability:dgu:hamil_dt} (resp. \eqref{eq:stability:pulkit:hamil_dt}), \eqref{eq:load_passivity:passivity_ineq}, and \eqref{eq:stability:line:hamil_dt}, the time derivative of \eqref{eq:stability:MG_hamil} is given by
\begin{align} \label{eq:stability:MG_hamil_dt}
    \dot{\Hamil}_\sMG(\xtilde_\sMG)=&
    \sum_{i \in \DD}\left(-  \psi_i(\xtilde_i)+ (\z_{i}-\z_i^*)^\Transpose ({\vd}_i-\bar{\vd}_i)\right) \nonumber\\
    + &\sum_{j \in \LL}\left(- \psi_j(\xtilde_j)+ ({\z_{j}-\bar{\z}_{j}})^\Transpose ({\vd}_j-\bar{\vd}_j)\right) \\
	+& \sum_{l \in \PP} \left(- \psi_l(\xtilde_l)+ ({\z_{l}-\bar{\z}_{l}})^\Transpose ({\vd}_l-\bar{\vd}^*_l)\right)\nonumber
\end{align}
As the ImG is an autonomous system with no more open interaction (coupling) ports, the supply rate of the ImG model is zero
\begin{align} \label{eq:stability:MG_supplyrate_zero}
    0=&
    \sum_{i \in \DD} ({\z_{i}-\z_{i}^*})^\Transpose ({\vd}_i-\bar{\vd}_i) \nonumber\\
    +& \sum_{j \in \LL} ({\z_{j}-\bar{\z}_{j}})^\Transpose ({\vd}_j-\bar{\vd}_j) \\
    +& \sum_{l \in \PP} ({\z_{l}-\bar{\z}_{l}})^\Transpose ({\vd}_l-\bar{\vd}^*_l) \nonumber
\end{align}
With \eqref{eq:stability:MG_supplyrate_zero}, $\psi_i(\xtilde_i)\geq0$, $\psi_j(\xtilde_j)\succ0$ under Proposition~\ref{prop:passivity_conditions:load}, and $\psi_l(\xtilde_l)\succ0$, \eqref{eq:stability:MG_hamil_dt} yields
\begin{align} \label{eq:stability:dHmg_dt_neg_semidef}
    \dot{\Hamil}_\sMG(\xtilde_\sMG) 
    \leq 0, \quad \forall \xtilde_\sMG \in \XX_\sMG,
\end{align}
which makes $\Hamil_\sMG(\xtilde_\sMG)$ a Lyapunov function for the equilibrium $\xtilde_\sMG=\x_\sMG-\xRef^*_\sMG=0$ and thus ensures stability of $\xRef^*_\sMG$. 
%
%
%
%
%
Furthermore, as $t \to \infty$, the state $\tilde{\x}_\sMG=\left[\xtilde_{1},\cdots,\xtilde_{L+D+P}\right]^T \in \XX_\sMG$ converges to the largest invariant set $M$ contained in
\begin{align}
\label{eq:stability:invariant_set1}
E= \left\{\xtilde_\sMG \in \XX_\sMG| \dot{\Hamil}_\sMG(\xtilde_\sMG) = 0 \right\}.
\end{align} 
From inserting \eqref{eq:stability:MG_supplyrate_zero} in \eqref{eq:stability:MG_hamil_dt}, we see that set $E$ in \eqref{eq:stability:invariant_set1} is characterized by
\begin{equation} \label{eq:stability:Hamil_dt_zero}
\psi_{i}(\xtilde_i)=0, \psi_{j}(\xtilde_j)=0, \psi_{l}(\xtilde_l)=0.
\end{equation}
%
With \eqref{eq:stability:MG_supplyrate_zero}, $\psi_i(\xtilde_i)\geq0$, $\psi_j(\xtilde_j)\succ0$ under Proposition~\ref{prop:passivity_conditions:load}, and $\psi_l(\xtilde_l)\succ0$, \eqref{eq:stability:Hamil_dt_zero} is satisfied by the family of vectors
\begin{equation}\label{eq:stability:family vectors}
\left\{
\begin{aligned}
\xtilde_i&=\!\!\begin{bmatrix}
\Zstate_{\sActuated 1,i} - \bar{\Zstate}_{\sActuated 1,i}\\ 
\Zstate_{\sActuated 2,i} - \bar{\Zstate}_{\sActuated 2,i}\\
\Zstate_{\sUnactuated 1,i} - \Zstate^*_{\sUnactuated 1,i}\\ 
\Zstate_{\sUnactuated 2,i} - \Zstate^*_{\sUnactuated 2,i}\\ 
\Zstate_{\sExtended1,i} -\bar{\Zstate}_{\sExtended1,i} \\
\Zstate_{\sExtended2,i} -\bar{\Zstate}_{\sExtended 2,i}
\end{bmatrix}\!\!=\!\!\begin{bmatrix}
0\\0\\ p_i\\ q_i\\r_i\\s_i
\end{bmatrix}\\
\xtilde_j&=\zero_2\\
\xtilde_l&=\zero_2
\end{aligned}
\right., i\in \DD_\sPHS, ~j\in \LL, ~l\in\PP,
\end{equation}
where $p_i\in\Reals$, $q_i\in\Reals$, $r_i\in\Reals$, and $s_i\in\Reals$. 
Now consider the evolution of a trajectory of \eqref{eq:stability:dgu:closed_loop_phs_xtilde:dynamics} starting from an initial point $\xtilde_i(0)\in E$. As ${\vd}_i-\bar{\vd}_i \propto \z_\sLine - \bar{\z}_\sLine=\frac{\xtilde_\sLine}{L_\sLine}, l \in \NN_i $ (cf.\ \eqref{eq:modelling:line_phs}, \eqref{eq:stability:inputDGULoad}), we obtain
\begin{subequations}\label{traj starting from P}
	\begin{align}
	\xtildedot_i(0) &=
    \left[ \J_{\sDesired\sZ,i}-\R_{\sDesired\sZ,i} \right] \frac{\partial \Hamil_{\sDesired \sZ,i}(\xtilde_i)}{\partial \xtilde_i}\big|_{\xtilde_i=\xtilde_i(0)} \nonumber\\
    &+\begin{bmatrix}
    \K\\
    \Zero_{2\times2}
    \end{bmatrix} ({\vd}_i-\bar{\vd}_i)\big|_{\xtilde_l=\zero_2}\\
	& = \left [ \begin{array}{c}
	\frac{-\nu_{11,i}p_i}{C_{ti}}\\ \frac{-\nu_{12,i}q_i}{C_{ti}}\\ \omega_0q_i-r_i \\ -\omega_0p_i-s_i\\ \frac{k_{\sI 1,i}p_i}{C_{ti}} \\ \frac{k_{\sI 2,i}q_i}{C_{ti}}
	\end{array} \right], \quad  i\in \DD_\sPHS.
	\end{align}
\end{subequations}
Then, in order to characterize the largest invariant set $M\subseteq E$, we impose that state trajectories of \eqref{traj starting from P} are confined in $E$ \eqref{eq:stability:family vectors} at any future time. This yields the set of equations 
\begin{equation}
\left\{\begin{array}{rl}
\frac{-\nu_{11,i}p_i}{\Cti}&=0\\ 
\frac{-\nu_{22,i}q_i}{\Cti}&=0\\ 
\omega_0q_i-r_i &=\dot{p}_i\\ 
-\omega_0p_i-s_i&=\dot{q}_i\\ 
\frac{k_{\sI 1,i}p_i}{\Cti}&=\dot{r}_i\\ 
\frac{k_{\sI 2,i}q_i}{\Cti}&=\dot{s}_i
\end{array} \right., \quad  i\in \DD_\sPHS,
\end{equation}
whose unique solution is 
\begin{equation}
p_i=q_i=r_i=s_i=0, \quad  i\in \DD_\sPHS.
\end{equation}
Furthermore, for DGUs $i\in\DD_\sPI$, it also holds that $\xtilde_i=\vec{0}$ and $\xtilde_i\in E, i\in\DD_\sPI$. The derivation is given in \cite{Nahata2021Thesis} and similar to the one provided in \cite{Tucci18TechRep}. 
Additionally, we note that an optional local load connected to the closed-loop DGU (see Remark~\ref{remark:dH_dt_load_at_dgu}) is a simplified special case of the analysis above. The load only induces additional damping such that $\psi_i(\xtilde_i)+\psi_{\RR_i}(\xtilde_i)\geq0$ in \eqref{eq:stability:Hamil_dt_zero}, which results in $\xtilde_i=\left[ 0, 0, 0, 0, r_i, s_i \right]^\Transpose$ in \eqref{eq:stability:family vectors}. 

Hence, the largest invariant set $M\subseteq E$ is the origin $\xtilde_\sMG =\zero_{2(L+3D+P)}$ implying that $\x_\sMG = \xRef^*_\sMG$. Thus, $\xRef^*_\sMG$ is asymptotically stable.
\IEEEQEDclosed
%
%
%
\begin{remark} \label{remark:controller_guidelines}
From the proof of Theorem~\ref{theorem:ImG_stability} we note that if the closed-loop DGUs were to be strictly EIP instead of merely EIP while retaining a positive definite storage function or Hamiltonian, respectively, \eqref{eq:stability:dHmg_dt_neg_semidef} would become
\begin{align} \label{eq:stability:dHmg_dt_neg_def}
    \dot{\Hamil}_\sMG(\xtilde_\sMG) 
    <0, \quad \forall \xtilde_\sMG \in \XX_\sMG,\, \xtilde_\sMG\neq \zero_{2(L+3D+P)}
\end{align}
directly implying asymptotic stability of $\xRef^*_\sMG$. This has the following implications for using DGUs controlled by other controllers $i\in\DD_\text{other}$ along the two analyzed in this work: (i) if the controller renders the closed-loop DGU strictly EIP with positive definite storage function, it can readily be used without endangering the asymptotic stability of $\xRef^*_\sMG$; (ii) if the closed-loop DGU is merely EIP, stability of $\xRef^*_\sMG$ is guaranteed in any case. To investigate whether asymptotic stability is maintained, only a separate analysis of the largest invariant set contained in  $\dot{\Hamil}_i(\xtilde_i) = 0, i \in \DD_\text{other} $ is necessary.
\end{remark}
\subsection{Stability Impacts of Actuator Saturation} \label{sec:stability:saturation}
The previous stability proof relies on the fact that the VSI of the DGU subsystem can set the voltage calculated by the control law \eqref{eq:control:overall_control_law}. However, in practice, power electronics are subject to physical constraints leading to limitations on the available control output $\Vtdq[i]$, i.e.\ \emph{actuator saturation}. As a consequence, the closed-loop system behavior changes and the resulting unknown equilibrium $\xRef_{i, \sSaturated} \neq \xRef^*_i$ leads to a new unknown ImG equilibrium $\xRef_{\sMG, \sSaturated}$ which may not be asymptotically stable anymore. 
In the sequel, we thus investigate the impact of actuator saturation on stability of the ImG equilibrium. For this, we drop the far-reaching assumption of an ideal three-phase AC voltage source in this section.

Firstly, without loss of generality, we make the following assumption:
\begin{assumption} \label{assumption:saturation}   
    The saturation limits of the VSI output voltage are $V_{\sInput i,\sSaturated}^\sD, V_{\sInput i,\sSaturated}^\sQ \in \Reals_{>0}$ yielding $V_{\sInput i}^\sD \in \left[ -V_{\sInput i,\sSaturated}^\sD, V_{\sInput i,\sSaturated}^\sD \right]$, $V_{\sInput i}^\sQ \in \left[ -V_{\sInput i,\sSaturated}^\sQ, V_{\sInput i,\sSaturated}^\sQ \right]$.
\end{assumption}
The VSI of a DGU $i \in \DD_\sSaturated \subseteq \DD$ is said to be \emph{saturated}, if the absolute value of the control law exceeds the saturation limits
\begin{equation}
    \begin{bmatrix}
        V_{\sInput i,\sSaturated}^\sD\\ 
        V_{\sInput i,\sSaturated}^\sQ
    \end{bmatrix}<
    \begin{bmatrix}
       | V_{\sInput i}^\sD |\\ 
        |V_{\sInput i}^\sQ |
    \end{bmatrix}, \quad i \in \DD_\sSaturated \subseteq \DD.
\end{equation}
Then, the DGU behaves like the open-loop system \eqref{eq:modelling:dgu_phs} with a constant control input. 
%
\begin{proposition} \label{prop:stability_AS}
Suppose Assumptions~\ref{assumption:balanced_three_phase}--\ref{assumption:saturation} hold.
Then, under actuator saturation with $\vu_i=\vuRef_i=\left[\pm V_{\sInput i,\sSaturated}^\sD, \pm V_{\sInput i,\sSaturated}^\sQ \right]^\Transpose$, the ISO-PHS \eqref{eq:modelling:dgu_phs} of the DGU $i \in \DD_\sSaturated \subseteq \DD$ is EIP with unknown equilibrium $\xRef_{i, \sSaturated}$ and the ImG exhibits an asymptotically stable equilibrium $\xRef_{\sMG, \sSaturated}$.
\end{proposition}
\IEEEproof
A key prerequisite for EIP is the existence of the unknown equilibrium with corresponding inputs and outputs. Under Assumption~\ref{assumption:higher_level_control}, $\xRef_{i, \sSaturated} \in \bar{\XX}_i$ is guaranteed, i.e.\ the unknown equilibrium under saturation exists and is inside the set of DGU operating points or equilibira, respectively, which is specified by higher level controls. Then, as the DGU ISO-PHS \eqref{eq:modelling:dgu_phs}  is linear with positive definite Hamiltonian and positive semidefinite dissipation matrix, EIP directly can be concluded from Lemma~\ref{lemma:eip_systems} (see also \cite[p.~136]{vdS17}). 
The shifted model is given by
\begin{subequations} \label{eq:stability:saturation:_phs_xtilde}
    \begin{align} \label{eq:stability:saturation:phs_xtilde:dynamics}
    \xtildedot_i &=
    \left[ \J_{i}-\R_{i} \right] \frac{\partial \Hamil_{i}(\xtilde_i)}{\partial \xtilde_i}+ \K_i \left({\vd}_i-\bar{\vd}_i\right)\\
	{\z_{i}-\bar{\z}_{i}} &= \K_i^\Transpose \frac{\partial \Hamil_{i}(\xtilde)}{\partial \xtilde}, \quad i\in\DD_\sSaturated\\
    \label{eq:stability:saturation:phs_xtilde:hamil}
     \Hamil_{i}(\xtilde_i) &= \frac{1}{2} \xtilde_{i}^\Transpose \Q_{i} \xtilde_{i}
\end{align}
\end{subequations}
with $\xtilde_i=\x_i-\xRef_{i,\sSaturated}$ and
\begin{equation} \label{eq:stability:staturated:psi}
    \psi_{i}(\xtilde_i)=\frac{\partial^\Transpose \Hamil_{i}(\xtilde_i)}{\partial \xtilde_i} \R_{i} \frac{\partial \Hamil_{i}(\xtilde_i)}{\partial \xtilde_i} \geq 0, \quad i\in\DD_\sSaturated
\end{equation}
(cf.\ \eqref{eq:stability:dgu:hamil_dt}).  Stability of $\xRef_{\sMG, \sSaturated}$ is thus ensured as per proof of Theorem  ~\ref{theorem:ImG_stability}. Asymptotic stability of $\xRef_{\sMG, \sSaturated}$ containing some $\xRef_{i, \sSaturated}, i \in \DD_\sSaturated \subseteq \DD$ follows by using LaSalle's Invariance principle as in Theorem~\ref{theorem:ImG_stability}. For this we note that $ \psi_{i}(\xtilde_i)=0$ in \eqref{eq:stability:staturated:psi} is satisfied by $\xtilde_i=\left[0, 0, p_i, q_i \right]^\Transpose, p_i, q_i \in \Reals$ and 
\begin{equation}
    	\xtildedot_i(0) =\begin{bmatrix}
    	   0\\
    	   0\\
    	   \dot{p}_i\\
    	   \dot{q}_i
    	\end{bmatrix}=\begin{bmatrix}
    	    \frac{-p_i}{\Cti}\\ 
            \frac{-q_i}{\Cti}\\ 
            \omega_0 q_i\\ 
            -\omega_0 p_i
    	\end{bmatrix},
\end{equation}
whose unique solution is $p_i=0, q_i=0$. \IEEEQEDclosed
\begin{remark}
Note that the results of Proposition~\ref{prop:stability_AS} are independent of the employed VSI controller, since the DGU simply behaves like the open-loop system \eqref{eq:modelling:dgu_phs} with constant control input.
\end{remark}
\section{Simulations} \label{sec:simulations}
In this section, we back up our theoretical findings by demonstrating asymptotic stability of \dq node voltage equilibira in an ImG comprising DGUs controlled with different controllers that render the closed-loop system EIP as well as ZIP and EXP loads operating standalone or connected to DGUs. We show how the decentralized, EIP-based controller design and decentralized online implementation provides plug-and-play functionality and allows for flexible changes in topology and loads without endangering stability. 


For this, we consider the CIGRE medium voltage distribution network benchmark \cite{Strunz14} operating at $\VampNominal=\SI{20}{\kilo\volt}$, $f_0=\SI{50}{\hertz}$ and implement Feeder 1 comprising 11 nodes as an ImG in {\sc Matlab}/{\sc Simulink} \emph{Simscape} (cf.\ Fig.~\ref{fig:CIGRE_microgrid}).
As illustrated in Fig.~\ref{fig:CIGRE_microgrid}, we connect DGUs at nodes $\DD=\{1,2,3,4,5,6 \}$ with $\DD_\sPI=\{1\}, \DD_\sPHS=\{2,3,4,5,6 \}$, ZIP loads at nodes $\LL_{\text{ZIP}}=\{1,2,4,5,7,8\}$, and EXP loads at nodes $\LL_{\text{EXP}}=\{9,10,11\}$.
%
DGU filters are parameterized identically with $\Rti=\SI{0.1}{\ohm}$, $\Lti=\SI{1.8}{\milli\henry}$, $\Cti=\SI{25}{\micro\farad}$. The control parameters are chosen as $\alpha_{11,i}=\alpha_{22,i}=-5$, $\nu_{11,i}=\nu_{22,i}=1$, $k_{\sI 1,i}=k_{\sI 2,i}=100 $ for $i\in\DD_\sPHS$ and $\KK_{11,i}=(K_{\text{P},i}\Lti+1)\One_{2\times2},K_{\text{P},i}=1000$,
\begin{equation*}
\renewcommand{\arraystretch}{0.8}
   \KK_{12,i}=\begin{bmatrix}
   \Rti             & -2 \freqRef \Lti \\
   2 \freqRef \Lti  & \Rti
   \end{bmatrix}-\KK_{13,i}\left(\frac{1}{K_{\text{P},i}}+1\right),
\end{equation*}
$\KK_{13,i}=K_{\text{I},i}\One_{2\times2}, K_{\text{I},i}=1000$ for $i\in \DD_\sPI$, respectively. The DGU references are set arbitrarily as in Table~\ref{table:simulation_dgu_refs}. 
The load parameters are given in Table~\ref{table:simulation_load_param} and all satisfy the EIP conditions of Proposition~\ref{prop:passivity_conditions:load}. The EXP load parameters are taken from \cite[pp.~44--45]{Farrokhabadi18techrep}, while we obtained the ZIP load parameters by equivalence transformation of the EXP loads\footnote{Cf.\ \cite[Eq. (38),(41)]{Farrokhabadi18techrep} with $Z^p=Z^q=\frac{5}{14}, I^p=I^q=\frac{1}{14}, P^p=P^q=\frac{1}{14}$.}.
The line parameters are given in Table~\ref{table:simulation_line_param} according to \cite[Tab.~6.12]{Strunz14} with the lengths as in Fig.~\ref{fig:CIGRE_microgrid}. 
%
In order to assess the compositional, plug-and-play functionality of the proposed EIP-based control framework, the simulations starts with DGU 5 disconnected from the ImG. At $t=\SI{4}{\second}$, it connects to the ImG.
To investigate robustness of the controlled ImG regarding load dynamics, at $t=\SI{5}{\second}$ and $t=\SI{6}{\second}$, loads 5 and 8, respectively, increase their overall demand $\lActNom[\sLoad], \lReacNom[\sLoad]$ by \SI{50}{\percent}, which results in new ZIP parameters (see Table~\ref{table:simulation_load_param} in brackets).
Additionally, the lines contain zero-sequence components and introduce parameter uncertainties to $\Cti$ by their parallel line capacitances. 

%
\begin{figure}
	\centering
    \includegraphics{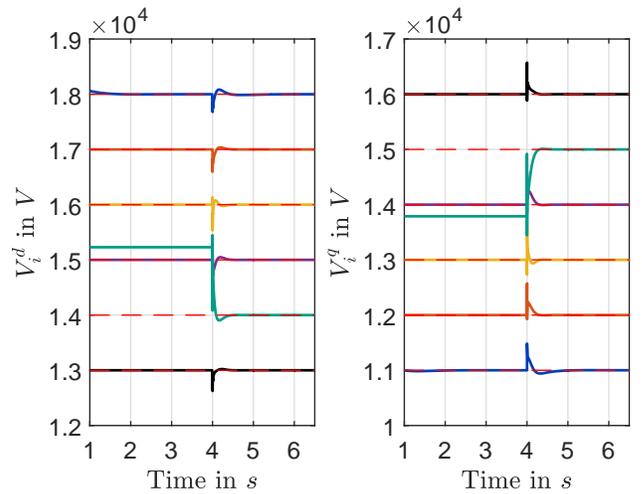}
	\caption{Voltages $\Vdq_i$ of the DGU nodes; DGU colors are given in Table~\ref{table:simulation_dgu_refs}}
	\label{fig:simulation_results_vdq}
\end{figure}
\begin{figure}
	\centering
    \includegraphics{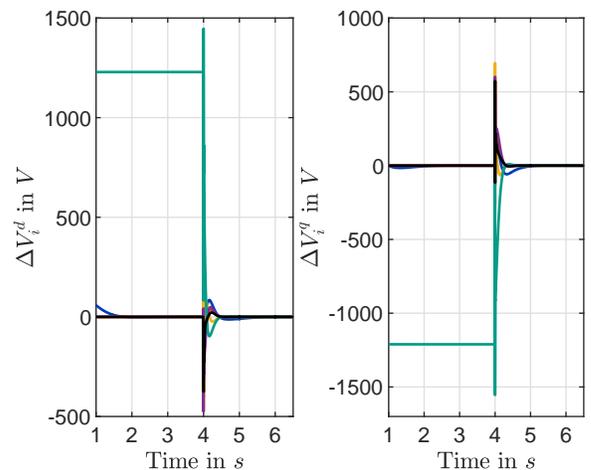}
	\caption{Voltage errors $\Delta \Vdq_i$ of the \dq DGU node voltages; DGU colors are given in Table~\ref{table:simulation_dgu_refs}}
	\label{fig:simulation_results_delta_vdq}
\end{figure}
\begin{figure}[h!]
	\centering
    \includegraphics{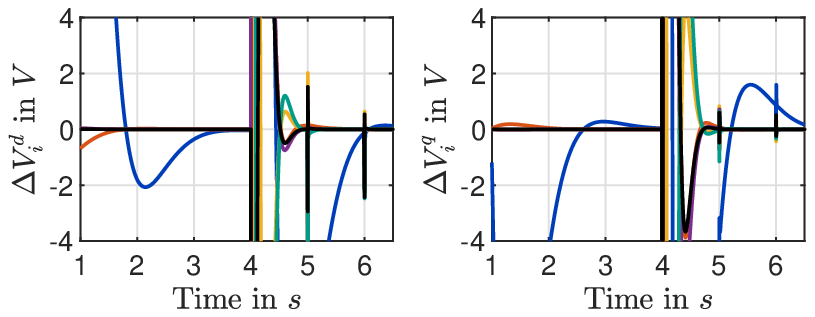}
	\caption{Zoomed in voltage errors $\Delta \Vdq_i$}
	\label{fig:simulation_results_delta_vdq_zoom}
\end{figure}
The results for the controlled DGU nodes are illustrated in Fig.~\ref{fig:simulation_results_vdq}, \ref{fig:simulation_results_delta_vdq}, and \ref{fig:simulation_results_delta_vdq_zoom}.
They show that $\VdqRef_i$ can be asymptotically stabilized throughout the complete simulation despite plug-and-play operations, unknown interaction currents, and parameter uncertainties. 
However, the voltage of the PI-controlled DGU $\DD_\sPI=\{1\}$ shows slower convergence behavior. After the start of the simulation and the connection of DGU 5, it takes around \SI{1.5}{\second} for its \dq voltage to settle within \SI{2}{\volt} (\SI{0.01}{\percent}) of the desired value.  By contrast, DGUs $\DD_\sPHS=\{2,3,4,5,6 \}$ achieves voltage errors below \SI{2}{\volt} (\SI{0.01}{\percent}) after around \SI{0.5}{\second}. Even under different tuning of $K_{\text{P},i},K_{\text{P},i}\in \Reals_{\geq 0}$ this behavior barely changes. Maximal voltage overshoots remain below an absolute value of \SI{700}{\volt} (\SI{3,5}{\percent})
The considerable \SI{50}{\percent} increases in load demands and the resulting load parameter changes at $t=\SI{5}{\second}$ and $t=\SI{6}{\second}$ cause very small voltage deviations in the order of several volts, i.e. around $\SI{0.1}{\percent}$, which are barely visible.
Note that while DGU 5 is disconnected, $\Vdq_5$ is not controlled and thus deviates clearly from the desired equilibrium.

Overall, the results demonstrate that EIP allows ImG subsystems to operate in a plug-and-play fashion without endangering ImG stability. Furthermore, integral action of the used controllers provides robustness against parameter uncertainties, unknown interaction currents, and load changes.
However, as expected, different control designs and implementations cause different closed-loop DGU behavior, which allows for stimulating complementary designs in order to address the manifold of control objectives in ImGs.


%
\begin{table}[!t]
	\centering
	\setlength{\tabcolsep}{4pt}
	\renewcommand{\arraystretch}{1.25}
	\caption{
		Reference Voltages with $\VampNominal=\SI{20}{\kilo\volt}$}
	\label{table:simulation_dgu_refs}
	\begin{tabular}{c c c c c c c}
	\noalign{\hrule height 1.0pt}
		DGU & 1  & 2 & 3  & 4  & 5  & 6 \\
		& \textbf{\color{colorDGU1}(blue)} & \textbf{\color{colorDGU2}(red)}  & \textbf{\color{colorDGU3}(yellow)} & \textbf{\color{colorDGU4}(purple)} & \textbf{\color{colorDGU5}(turquoise)} & \textbf{\color{colorDGU6}(black)} \\
		\hline
		$\VdRef (\si{\pu})$ &   0.9  &      0.85    &   0.8     &   0.75    &   0.7     &   0.65 \\ 
		$\VqRef (\si{\pu})$ &   0.55 &      0.6     &   0.65    &   0.7     &   0.75    &   0.8   \\
		\noalign{\hrule height 1.0pt}
	\end{tabular}
\end{table}
%
%
\begin{table}[!t]
	\centering
	\renewcommand{\arraystretch}{1.25}
	\caption{
		ZIP and EXP Load Parameters}
	\label{table:simulation_load_param}
	\begin{tabular}{c c c c c c}
	\noalign{\hrule height 1.0pt}
        $\sLoad$ & $\lYp[\sLoad]$ (\si{\micro\siemens})  & $\lIp[\sLoad]$ (\si{\ampere}) & $\lPp[\sLoad]$ (\si{\kilo\watt}) & $\lActNom[\sLoad]$ (\si{\kilo\watt}) & $\lNp[\sLoad]$\\
        & $\lYq[\sLoad]$  (\si{\micro\siemens})  & $\lIq[\sLoad]$ (\si{\ampere}) & $\lPq[\sLoad]$ (\si{\kilo\var}) & $\lReacNom[\sLoad]$ (\si{\kilo\var}) &  $\lNq[\sLoad]$\\
		\hline
		1   & 75    & 0.3   & 6     & 84    & 1.57 \\
		2   & 97.5  & 0.39  & 7.8   & 109.2 & 1.57\\
		4   & 51.96 & 0.208 & 4.157 & 58.2  & 1.57\\
		5   & 80.8  & 0.323 & 6.464 & 90.5  & 1.57 \\
		    & (121.2) & (0.485) & (9.696) & (135.75) & (1.57) \\
		7   & 51.96 & 0.208 & 4.157 & 58.2  & 1.57 \\
		8   & 58.48 & 0.234 & 4.679 & 65.5  & 1.57 \\
		    & (87.72) & (0.351) & (7.018) & (98.25) & (1.57) \\
		\hdashline
		9   &       &       &       & 50.9  & 1.5 \\
		10  &       &       &       & 72.8  & 1.5 \\
		11  &       &       &       & 80    & 1.5 \\
		\noalign{\hrule height 1.0pt}
	\end{tabular}
\end{table}
%
%
\begin{table}[!t]
	\centering
	\renewcommand{\arraystretch}{1.25}
	\caption{
		Electrical Line Parameters}
	\label{table:simulation_line_param}
	\begin{tabular}{l c c c c}
		\noalign{\hrule height 1.0pt}
		&&& Positive-sequence & Zero-sequence \\
		\hline
		Resistance & $R_\sLine$ & (\si{\ohm \per \km})  & 0.343 & 0.817 \\
		Inductance & $L_\sLine$ & (\si{\milli\henry \per \km})  & 0.875 & 5.087 \\
		Capacitance & $C_\sLine$ & (\si{\nano\farad \per \km}) & 151.7  & 151.7 \\
		\noalign{\hrule height 1.0pt}
	\end{tabular}
\end{table}
\section{Conclusion}
In this work, we delineated a unified control framework which ensures voltage and frequency stability within an ImG network, i.e.\ stability of the unknown ImG equilibrium, based on the EIP properties of the ImG subsystems---dynamic RLC lines, nonlinear ZIP and EXP loads, and DGUs controlled with different controllers.
This gives rise to a modular and scalable ImG network wherein multiple subsystems, if EIP, can seamlessly enter or leave without spoiling ImG stability.
For the DGUs, we then devised decentralized controllers based on a port-Hamitonian design (IDA-PBC with additional IA), which render the closed-loop DGU EIP.
By including the passivating DGU controllers from \cite{Nahata19ecc} in our subsequent stability analysis, we demonstrated the compositional, unifying nature provided by the proposed EIP control framework in which ImG operators are not restricted to the controllers proposed in this work.
Indeed, the insights obtained from the proof of asymptotic voltage and frequency stability provide very general guidelines for ImG operators to employ other decentralized DGU controllers, which allows for fruitful competition and collaboration.
Furthermore, we showed that the practically very relevant issue of actuator saturation does not undermine the asymptotic stability of the ImG equilibrium, independent of the employed controller.
\end{document}